\documentclass[12pt]{article}
\usepackage[a4paper,top=3cm,bottom=3cm,left=3cm,right=3cm,marginparwidth=1.75cm]{geometry}
\usepackage{amsmath,amsfonts,amssymb,bm,mathrsfs,mathtools,overpic,amsthm,subfig,svg}
\usepackage{graphicx,tikz}
\usetikzlibrary{arrows,decorations.markings}
\usepackage[english]{babel}
\usepackage{authblk}

\usepackage[authoryear,square]{natbib}
\graphicspath{{./}{./figs}}

\definecolor{myred}{RGB}{200,0,0}
\colorlet{ColorPink}{red!20}

\usepackage{tikz,tikz-3dplot}
\tikzset{
  font={\footnotesize}}
\usetikzlibrary{math,positioning,calc}
\usetikzlibrary{decorations.pathreplacing}
\usepackage{pgfplots}

\usepackage[colorinlistoftodos]{todonotes}
\usepackage[colorlinks=true, allcolors=blue]{hyperref}

\def\d{{\mathrm d}}

\newcommand{\reseteqnos}[1]{\renewcommand{\theequation}{#1.\arabic{equation}}
\setcounter{equation}{0}}

\newcommand{\bea}{\begin{eqnarray}}
\newcommand{\eea}{\end{eqnarray}}
\newcommand{\be}{\begin{equation}}
\newcommand{\ee}{\end{equation}}
\newcommand{\bes}{\begin{subequations}}
\newcommand{\ees}{\end{subequations}}
\def\vecu{{\text{\boldmath$u$}}}

\def\vect{{\text{\boldmath$t$}}}
\def\vecn{{\text{\boldmath$n$}}}
\def\ha{{\hat{a}}}
\def\hx{{\hat{x}}}
\def\hX{{\hat{X}}}
\def\hz{{\hat{z}}}
\def\hZ{{\hat{Z}}}
\def\hu{{\hat{u}}}

\def\hp{{\hat{p}}}
\def\hm{{\hat{m}}}
\def\hk{{\hat{k}}}
\def\hh{{\hat{h}}}
\def\hL{{\hat{L}}}
\def\hR{{\hat{R}}}
\def\hkappa{{\hat{\kappa}}}
\def\hdel{{\hat{\delta}}}
\def\hbeta{{\hat{\beta}}}
\def\hOm{{\hat{\Omega}}}
\def\hmu{{\hat{\mu}}}
\def\pd{{\partial}}
\def\d{{\mbox{d}}}
\def\ie{i.\,e.\ }

\title{Optimising the flow through a concertinaed filtration membrane}
\author[1]{Victoria E. Pereira}
\author[1]{Mohit P. Dalwadi}
\author[2]{Enrique Ruiz-Trejo}
\author[1]{Ian M. Griffiths}
\affil[1]{Mathematical Institute, University of Oxford, Oxford OX2 6GG, UK}
\affil[2]{Smart Separations Ltd., 40 Occam Rd, Guildford GU2 7YG, UK}

\begin{document}

\maketitle

\begin{abstract}
Membrane filtration is a vital industrial process, with applications including air purification and blood filtration. In this paper, we study the optimal design for a concertinaed filtration membrane composed of angled porous membranes and dead-ends. We examine how the filter performance depends on the angle, position, thickness, and permeance of the membrane, through a combination of numerical and asymptotic approaches, the latter in the limit of a slightly angled membrane. We find that, for a membrane of fixed angle and physical properties, there can exist multiple membrane positions that maximise the flux for an applied pressure difference. More generally, we show that while the maximal flux achievable depends on the membrane thickness and permeance, the optimal membrane configuration is always in one of two setups: centred and diagonal across the full domain; or angled and in the corner of the domain.
\end{abstract}

\section{Introduction}\reseteqnos{1}

Membrane filtration is a process used for the clarification, purification, and separation of fluid mixtures \citep{noble1995membrane}. This field has many important applications from filtering of blood to purifying water and air \citep{van2007bioprocess, lee2011review}. In a typical filtration system, a mixture of fluid and contaminant particles is passed through a porous membrane; the fluid passes through while the particles are retained, either on the surface of the membrane or within the membrane structure. 

Filtration membranes can be divided broadly into two main types: \emph{fibre-based membranes} comprise a random network of fibres, and \emph{membrane surfaces} are manufactured with pores. While fibre-based membranes have historically been easier to manufacture, they come with several drawbacks. For example, there is minimal control of their pore structure and they have a large environmental footprint since the fibres clog and have to be discarded, increasing waste to landfill. Furthermore, new techniques are emerging to manufacture such membranes more easily \citep{SSL}. As such, the second type of manufactured membranes, which have the benefit of a controlled pore structure and the potential to be cleaned and reused, are often the preferable choice. Since this type of membrane can be precisely manufactured, it is important to understand how to maximise the flux through such filters through design parameters.

As the unwanted particles are filtered out of the fluid mixture, the membrane blocks over time. There are two main blocking mechanisms: \textit{caking} is the build-up of large particles on the surface of the membrane, and \textit{internal blocking} is the clogging inside the actual pore. Blocking is inevitable in filtration. It affects the flux of fluid through the membrane, and consequently the efficiency of a filter is strongly coupled to any blocking.

The blocking of filters motivates the study of filtration devices, with the goal of finding the optimal filter design, \ie the design that maximises flux through the filter and minimises the effect of blocking. Filtration devices can be classified by the direction of flow: fluid mixtures are passed through filtration membranes with the flow either perpendicular to the membrane surface in what is called \textit{dead-end flow}, or parallel to the surface in \textit{crossflow} \citep{noble1995membrane}. 

When using a filter in dead-end flow, all of the components of the fluid mixture are either passed through or retained by the membrane. While dead-end flow is simple to use, the technique is unsuitable for processing high volumes as the normal flow results in cake layers building up quickly. Crossflow is motivated by the end of minimising the caking on the membrane surface; the parallel flow exposes less of the fluid mixture to the membrane, therefore processing the fluid mixture slower, but continuously `washing' the membrane, which inhibits cake build up. Thus crossflow is better for filtering high volumes of fluid. However this technique also has its disadvantages as the process requires recirculation of the fluid, and is therefore more complex and energy consuming.

\textit{Direct-flow} filtration devices consist of stacked crossflow-membranes with capped ends therefore utilising both the dead-end and crossflow techniques. Direct-flow filters are more economical and more energy efficient than crossflow devices, while still benefiting from the reduced cake build-up. Typically, direct-flow devices comprise vertically stacked filtration membranes. 

There have been numerous studies into the mathematical modelling of vertically stacked direct-flow filtration devices \citep{herterich2017optimizing,xu2017pressure,wang2017}. Of particular relevance to this paper is \cite{herterich2017optimizing} who developed a mathematical model for flow through a direct-flow device comprising a system of stacked cylinders, where the porous cylinder walls provide the parallel membrane surfaces for crossflow. A key observation was that when the membranes were stacked too closely or too sparsely, the total flux through the device was reduced and hence there was an optimal stacking distance.

\begin{figure}
\centering
\subfloat[][\label{fig1a}]{
\hspace*{-30pt}
\raisebox{-10pt}{
\begin{tikzpicture}
    \node[anchor=south west,inner sep=0] at (0,0) {\includegraphics[scale=0.55,tics=10]{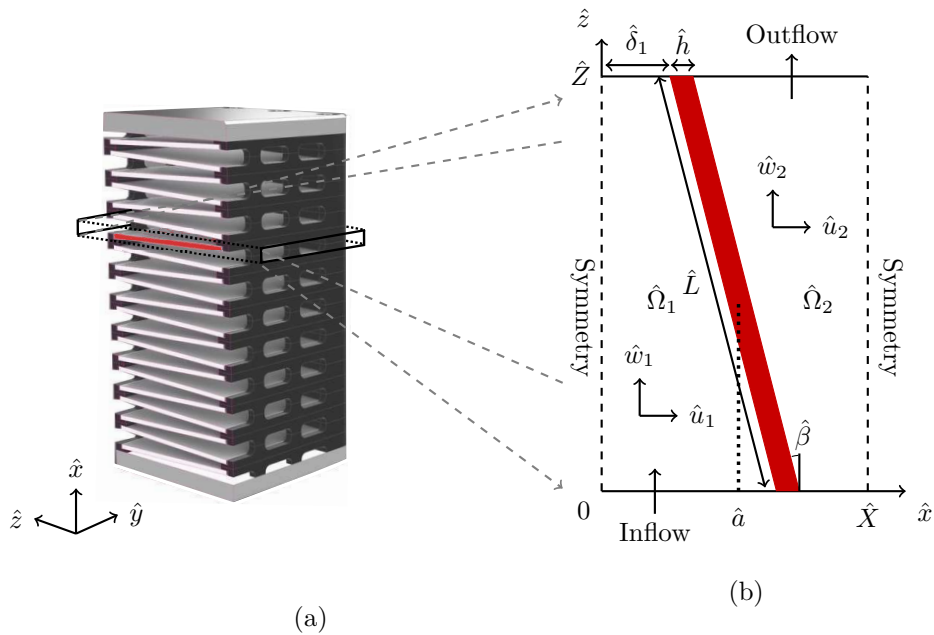}};
    \draw [thick,dashed,gray,->] (2.45,4.35) -- (6.4,1.4);
    \draw [thick,dashed,gray] (2.45,4.55) -- (6.4,2.8);
    \draw [thick,dashed,gray,->] (0.05,4.75) -- (6.4,6.57);
    \draw [thick,dashed,gray] (0.04,4.95) -- (6.4,6);
    \draw [thick,->] (0,0.8) -- (0,1.4) node[anchor=south] {$\hat{x}$};
    \draw [thick,->] (0,0.8) -- (-0.55,1) node[anchor=east] {$\hat{z}$};
    \draw [thick,->] (0,0.8) -- (0.55,1.05) node[anchor=west] {$\hat{y}$};
\end{tikzpicture}
}}
\hspace*{-17pt}
\subfloat[][\label{fig1b}]{
\raisebox{6pt}{
\begin{tikzpicture}[remember picture]
\draw [thick,->] (0,0) -- (4,0) node[anchor=north west] {$\hat{x}$};
\draw [thick,->] (0,5.5) -- (0,6) node[anchor=south east] {$\hat{z}$};
\draw [thick,dashed] (0,0) -- (0, 5.5);
\draw [thick,dashed] (3.5,0) -- (3.5, 5.5);
\draw [thick] (0,5.5) -- (3.5, 5.5);
\node [below left] at (0,0) {0};
\node [below] at (3.5,0) {$\hat{X}$};
\node [left] at (0,5.55) {$\hat{Z}$};
\node [rotate=-90] at (3.8,2.3) {Symmetry};   
\node [rotate=-90] at (-0.25,2.3) {Symmetry};
\node [above] at (2.5,5.8) {Outflow};
\node [above] at (0.7,-0.8) {Inflow};
\draw [thick,->] (0.7,-0.3) -- (0.7,0.3);
\draw [thick,->] (2.5,5.2) -- (2.5,5.8);
\draw [myred, fill=myred, opacity=1] (2.6,0) -- (1.2,5.5) -- (0.9,5.5) -- (2.3,0);
\draw[arrows=<->,line width=0.8pt](0.75,5.5)--(2.15,0.05) node[midway,anchor=east] {$\hat{L}$};
\draw[line width=1.3pt,dotted] (1.8,0) -- (1.8, 2.5);
\node [below] at (1.8, -0.07) {$\hat{a}$};
\draw[arrows=<->,line width=0.8pt](0.05,5.65)--(0.89,5.65) node[midway,anchor=south] {$\hat{\delta}_1$};
\draw[arrows=<->,line width=0.8pt](0.91,5.65)--(1.23,5.65) node[midway,anchor=south] {$\hat{h}$};
\draw [thick] (2.6,0) -- (2.6,0.5);
\draw (2.5,0.46) arc (105:95:0.5);
\node [above] at (2.65,0.4) {$\hat{\beta}$};
\node at (0.8, 2.55) {$\hat{\Omega}_1$};
\node at (2.85, 2.55) {$\hat{\Omega}_{2}$};
\draw [thick,->] (0.5,1) -- (0.5,1.5) node[anchor=south] {$\hat{w}_1$};
\draw [thick,->] (0.5,1) -- (1,1) node[anchor=west] {$\hat{u}_1$};
\draw [thick,->] (2.25,3.5) -- (2.25,4) node[anchor=south] {$\hat{w}_2$};
\draw [thick,->] (2.25,3.5) -- (2.75,3.5) node[anchor=west] {$\hat{u}_2$}; 
\end{tikzpicture}
}}
\caption{Schematic of the concertinaed filtration device. Figure~\ref{fig1a} shows the full three-dimensional device; reproduced from \cite{SSL} with permission. Figure~\ref{fig1b} depicts the two-dimensional domain of a single filtration module.}
\label{fig1}
\end{figure}

In this paper we will examine how the design setup affects the flow through a direct-flow device with angled membranes. An industrial example and motivation for this is the direct-flow device designed by \cite{SSL}. A schematic for this filter is shown in figure~\ref{fig1a}. There are similarities between a direct-flow device with angled membranes and a pleated filtration membrane. The main distinction between the two is that fluid can travel through the full porous pleated membrane structure, whereas direct-flow devices comprise  dead ends (\ie $\hat{\delta}_1 >0$ in figure~\ref{fig1b}) through which the flow cannot travel. The dead ends change the flow structure within the filter. There has been research studying the flow through pleated membranes to determine the optimal pleat density in such a membrane \citep{chen1995optimization, rebai2010semi}. Additionally, previous studies have examined the effect of the geometry on the pressure drop achieved across a single pleat for a given flux \citep{caesar2002influence, saleh2016analytical,theron2017numerical}.

Recent relevant studies by \cite{sanaei2016flow} and \cite{sun2019modeling} model the flow and blocking mechanisms in pleated devices, adopting the assumption of Darcy flow throughout the whole domain composed of rectangular pleats. The focus of these works was to find the filter design that optimised the life-time of the filter accounting for transient blocking dynamics. Earlier work by \cite{king1996asymptotic} also models the flow through a pleated filtration membrane. The authors account for Stokes flow between the membranes driven by a prescribed flux through rectangular pleats with porous walls and ends, with a focus on the shape of cake build-up along the membrane surface. The distinction between these past works examining the transient flow through a pleated device and the work presented here is that our focus is on determining optimal experimental setups to maximise the steady flux through a direct-flow device.

\subsection{Problem statement}

In this paper, we present a comprehensive study of a direct-flow device with angled membranes. Specifically, we determine how the geometric and operating parameters affect the flow through the device. Our main objective is to determine the optimal setup to maximise the flux for a given pressure drop. We develop a mathematical model of flow through a direct-flow filter comprising angled membranes within the filter structure. The angled membranes increase the membrane surface area in a concertinaed filtration device, as shown in figure~\ref{fig1a}.

We consider a filtration device with openings through which the fluid is introduced and removed, closed ends, and porous membranes. The concertina-structure comprises repeated single modules, as shown in figure~\ref{fig1a}. The flow is driven by a pressure gradient, and we study how the optimal flow depends on the system parameters defining a single repeated module. 

We build on the modelling approach used by \cite{herterich2017optimizing}, who considered the similar problem of an applied flux through vertical walls, to study the design of an angled membrane module that maximises the flux for a given applied pressure difference. The structure of this paper is as follows. In Section~\ref{sec:model} we formulate the flow problem through a single repeated module. In Section~\ref{sec:vary_a_beta}, we consider an infinitely thin membrane of specified permeance and angle, and determine the membrane position within the module that results in the maximum flux. We analyse two parameter regimes: the first for a slightly angled membrane in Section~\ref{sec:asymptotics}, and the second for an arbitrarily angled membrane in Section~\ref{sec:numerics_k1}. In Section~\ref{sec:vary_k_h}, we determine the optimal angle and position associated to the maximum flux through a membrane of specified thickness and permeance. We first consider vanishingly thin membranes and study the role of permeance in Section~\ref{sec:numerics_varyk}. We then extend our study to membranes of finite thickness in Section~\ref{sec:numerics_varyh}. In Section~\ref{sec:conc} we discuss the implications and conclusions of our work, as well as avenues of future work.

\section{Model development}\label{sec:model}
\reseteqnos{2}

We model the flow through a single repeated module of the filtration device shown in figure~\ref{fig1}.
The system geometry is such that we can assume variations in the $\hat{y}$-direction to be small. We therefore focus our attention to the two-dimensional domain in $(\hx,\hz)$-space depicted in figure~\ref{fig1b}. We denote dimensional and dimensionless quantities with and without hats, respectively. The position of the membrane is described by the left membrane wall $\hx=\hm(\hz)$ given by
\be \label{2.0}\hm(\hz) = \ha + \tfrac{1}{2}\left(\hL\sin\hbeta - \hh\right) -\hz \tan\hbeta,\ee
where $\hL$ the length of the membrane, $\hh$ is the membrane thickness, $\ha$ is the midpoint, and $\hbeta$ is the angle of the membrane; see figure~\ref{fig1b}. The midpoint of the membrane, $\ha = \hdel_1 + (\hL\sin\hbeta+\hh)/2$, specifies the distance between the centres of the membranes in neighbouring modules. 

The flow domain comprises two subdomains: 
\bes \label{2.1}
\bea
\hOm_1 &=& \{ \hx\in[0,\hm(\hz)],\, \hz\in[0,\hZ]\},\\[0.2em]
\hOm_{2} &=& \{ \hx \in [\hm(\hz)+\hh, \hX],\, \hz\in[0,\hZ]\},
\eea
\ees
where $\hZ = \hL\cos\hbeta$ and $\hX$ specify the vertical and horizontal lengths of the domain, respectively. Note that while $\hX$ and $\hZ$ are specified constants, the membrane length $\hL$ and the prescribed angle $\hbeta$ are coupled. In this paper, we refer to the prescribed angle $\hbeta$ and this will constitute a key experimental parameter that can be varied. Moreover, without loss of generality, we consider $\hbeta > 0$ and note that neighbouring modules have $\hbeta < 0$ with the subdomains switched for symmetric flux, as can be seen in figure~\ref{fig1a}.

The flow enters $\hOm_1$ at $\hz=0$ with constant inlet pressure $\hp_{\text{in}}$, and exits $\hOm_2$ at $\hz=1$ with constant outlet pressure $\hp_{\text{out}}$. The pressure difference $(\hp_{\text{in}}-\hp_{\text{out}})>0$ drives the flow through the porous membrane, and a particular quantity of interest is the flux through the domain. 

\begin{table}
\centering
\def\arraystretch{1.3}
\begin{tabular}{|c|c|c| } 
 \hline
 Parameter &  Value \\
 \hline 
Vertical height, $\hZ$ & 50 mm \\
Horizontal length, $\hX$ & 3.6 mm \\
Membrane thickness, $\hh$ & 1.2 mm \\
Air viscosity, $\hmu$ & $1.81\times 10^{-5}$ Pa s\\
Air density, $\hat{\rho}$ & 1.2 kg m$^{-3}$\\
Vertical velocity scale, $\hat{W}_0$ & 50 mm s$^{-1}$\\
\hline
 \end{tabular}
 \caption{Parameter values for air flow through the filtration device shown in figure~\ref{fig1}; values for $\hZ$, $\hX$, $\hh$, and $\hat{W}_0$ provided by \cite{SSL}.}
 \label{tab:parameters}
\end{table}

\subsection{Governing equations}

The ratio between the horizontal and vertical domain lengths is $\epsilon = \hX /\hZ$. Using the parameter values given in table \ref{tab:parameters}, we find that $\epsilon \ll 1$. Moreover, the Reynolds number Re =$\hat{\rho}\hat{W}_0 \hZ/\hmu$ and the values in table \ref{tab:parameters} yield a small reduced Reynolds number, $\epsilon^2$Re. We therefore use the Stokes flow equations to describe the flow in $\hat{\Omega}_1$ and $\hat{\Omega}_2$:
\bes \label{2.2}
\bea 
\hat{\mu} \hat{\bm\nabla}^2 \hat{\vecu} - \hat{\bm\nabla} \hp &=& 0,\\
\hat{\bm\nabla} \cdot \hat{\vecu} &=& 0,
\eea
\ees
where $\hat{\vecu}(\hx,\hz)=(\hu(\hx,\hz),\hat{w}(\hx,\hz))$ is the velocity, $\hp$ is the pressure, and $\hat{\mu}$ is the constant fluid viscosity. We seek steady solutions in both subdomains denoting the variables in $\hOm_1$ by $(\hu_1,\hat{w}_1,\hp_1)$ and those in $\hOm_{2}$ by $(\hu_{2},\hat{w}_{2},\hp_{2})$.

As described above, the flow enters the domain along $\hz=0$ into $\hOm_1$, passes through the membrane at $\hx =\hm(\hz)$ and exits along $\hz=\hZ$ from $\hOm_{2}$, driven by a pressure gradient. The boundary conditions at the inlet and outlet are
\bes \label{2.3}\bea
\hp_1 &=& \hp_{\text{in}} \quad\hspace*{4pt} \mbox{ at } \hz=0,\hspace*{7pt} \hx\in[0,\hm(0)],\\
\hp_{2} &=&  \hp_{\text{out}} \quad\mbox{ at } \hz=\hZ,\,\, \hx\in[\hm(\hZ)+\hh, \hX].
\eea \ees
There are closed ends at $\hz=\hZ$ in $\hOm_{1}$ and at $\hz=0$ at $\hOm_{2}$ through which the flow cannot penetrate. The following boundary conditions enforce no penetration as well as no slip:
\bes \label{2.4}\bea
\hat{\vecu}_1 &=& \textbf{0} \quad \mbox{ at } \hz=\hZ,\hspace*{7pt} \hx\in[0,\hm(\hZ)],\\
\hat{\vecu}_{2} &=& \textbf{0} \quad \mbox{ at } \hz=0,\hspace*{9pt} \hx\in[\hm(0)+\hh, \hX].
\eea \ees
Since the entire filter consists of a periodic array of modules, we impose symmetry conditions across $\hx=0$ and $\hx=\hX$, yielding
\bes \label{2.5}\bea
\hu_1 &=& 0, \quad \frac{\pd \hat{w}_1}{\pd \hx} =\, 0 \quad \mbox{ at } \hx=0,\\[0.5em]
\hu_{2} &=&  0, \quad \frac{\pd \hat{w}_{2}}{\pd \hx} =\, 0 \quad \mbox{ at } \hx=\hX.
\eea \ees

The normal flow through the filter is modelled as a porous flow (\ie  governed by the pressure difference and permeance). Thus, employing Darcy flow across the membrane provides the boundary condition for the velocity in $\hOm_1$:
\be \label{2.7}
\hat{\vecu}_1\cdot \vecn = \hkappa \left[\hp_1(\hm,\hz) - \hp_{2}(\hm+\hh,\hz)\right] \quad \mbox{ at } \hx=\hm(\hz),
\ee
where $\vecn=(\cos\hbeta, \sin\hbeta)$ is the unit normal vector to the membrane and $\hkappa =  \hk/\hat{\mu} \hh$ is the membrane permeance, where $\hk$ is the permeability of the membrane. Note that the flow resistance due to the membrane is $\hR = 1/\hkappa$.

A flux balance through the membrane provides the corresponding boundary condition for the velocity in $\hOm_{2}$:
\be \label{2.8}
\hat{\vecu}_{1}|_{\hx=\hm} = \hat{\vecu}_{2}|_{\hx=\hm+\hh}.
\ee

Finally, we need conditions at the permeable membrane layer. While \cite{beavers1967} provide an appropriate tangential slip-flow boundary condition, \cite{griffiths2013} show that including slip does not have a significant effect on the flow. Hence, for simplicity, we impose no slip on both sides of the membrane:
\bes\label{2.6}\bea 
\hat{\vecu}_1\cdot \vect &=& 0 \quad \mbox{ at } \hx=\hm(\hz),\\
\hat{\vecu}_{2}\cdot \vect &=& 0 \quad \mbox{ at } \hx=\hm(\hz)+\hh,
\eea \ees
where $\vect=(-\sin\hbeta, \cos\hbeta)$ is the unit tangent vector to the membrane.

\subsection{Dimensionless model}

We non-dimensionalise the system \eqref{2.2}--\eqref{2.8} by introducing the following scalings:
\be \label{2.9}
\hz = \hZ z, \quad \hx = \epsilon \hZ x, \quad \hbeta = \epsilon \beta, \quad \hu = \epsilon \hat{W}_0 u, \quad \hat{w} = \hat{W}_0 w, \quad \hp = \dfrac{\hmu \hat{W}_0}{\epsilon^2 \hZ} p+\hp_{\text{out}};
\ee
recalling that $\epsilon = \hX/\hZ$. Applying \eqref{2.9} to \eqref{2.2} and taking the limit $\epsilon\to 0$ yields 


\refstepcounter{equation}
\[
p_x = 0, \quad w_{xx} - p_z = 0, \quad  u_x + w_z = 0. 
\eqno{(\theequation{\mathit{a}-\mathit{c}})}\label{2.12}
\]

\noindent Hence, in both subdomains the lubrication equations govern the flow.

The leading-order position of the membrane \eqref{2.0} becomes:
\be \label{2.10}
m(z) = a - \tfrac{1}{2}h + \beta\left(\tfrac{1}{2}-z\right),
\ee
where $m=\hm/\hX$, $a=\ha/\hX$, $h=\hh/\hX$, and the dimensionless membrane angle is $\beta \in [0,1]$. The corresponding dimensionless subdomains from \eqref{2.1} are given by
\bes
\bea
\Omega_1 &=& \{ x\in[0,m(z)],\, z\in[0,1]\},\\
\Omega_{2} &=& \{ x \in [m(z)+h, 1],\, z\in[0,1]\}.
\eea
\ees
The dimensionless domain is identical to that shown in figure~\ref{fig1b} with the hats removed and with $X=Z=1$. Recall that while the domain size specified by $Z$ and $X$ is fixed, the membrane length varies with $\beta$. 

Applying the scaling \eqref{2.9} to the boundary conditions prescribed above, the pressure conditions \eqref{2.3} become
\bes \bea
p_1 &=& 1 \quad\hspace*{4pt} \mbox{ at } z=0,\hspace*{7pt} x\in[0,m(0)],\label{2.13a}\\
p_{2} &=&  0 \quad\hspace*{4pt} \mbox{ at } z=1,\hspace*{7pt} x\in[m(1)+h, 1],\label{2.13b}
\eea \ees
and the boundary conditions at the closed walls \eqref{2.4} become
\bes \label{2.14}\bea
w_1 &=& 0 \quad \mbox{ at } z=1,\hspace*{7pt} x\in[0,m(1)],\label{2.14a}\\
w_{2} &=&  0 \quad \mbox{ at } z=0,\hspace*{7pt} x\in[m(0)+h, 1].\label{2.14b}
\eea \ees
We note that information regarding the tangential velocity on the closed walls in \eqref{2.14} is lost since we are working in the lubrication regime. The dimensionless symmetry conditions from \eqref{2.5} are
\bes \label{2.15}\bea
u_1 &=& 0, \quad \frac{\pd w_1}{\pd x} \,\,=\, 0 \quad \mbox{ at } x=0,\label{2.15a}\\[0.5em]
u_{2} &=&  0, \quad \frac{\pd w_{2}}{\pd x} =\, 0\,\, \quad \mbox{ at } x = 1.\label{2.15b}
\eea \ees

We define the dimensionless permeance by $\kappa = \hmu \hkappa/\epsilon^3 \hZ$, and the dimensionless permeability by $k = \hk/\epsilon^3 \hZ \hX$. With these definitions, the dimensionless permeance is \mbox{$\kappa = k/h$}. The Darcy flow condition through the membrane \eqref{2.7} becomes
 \be \label{2.17}
u_1 =\kappa \left[p_1(m,z) - p_{2}(m(z)+h,z)\right] \quad \mbox{ at } x=m(z),
\ee
where the dimensionless membrane resistance is $R=1/\kappa$.
The flux-balance boundary condition across the membrane \eqref{2.8} becomes
\be\label{2.18}
u_{1}|_{x=m} = u_{2}|_{x=m+h}.
\ee
Finally, the no-slip condition \eqref{2.6}, exploiting the small-angle of the membrane, is 
\bes \label{2.16}  \bea 
w_1 &=& 0 \quad \mbox{ at } x=m(z),\label{2.16a}\\
w_{2} &=& 0 \quad \mbox{ at } x=m(z)+h.\label{2.16b}
\eea \ees

\subsection{Model reduction}

Since $p_1=p_1(z)$ and $p_2=p_2(z)$ from (\ref{2.12}{\color{red}\textit{a}}), we can reduce the full problem \eqref{2.12}--\eqref{2.16} to modified Reynolds equations. Since we have two coupled domains, we derive two coupled second-order ordinary differential equations (ODEs)  for the pressure; one ODE for each subdomain. The details are as follows.

In the region $\Omega_1$ we solve (\ref{2.12}{\color{red}\textit{b}}) and (\ref{2.12}{\color{red}\textit{c}}) with the boundary conditions \eqref{2.15a} and \eqref{2.16a} to find the following relationships between the velocities and the pressure:
\bes\label{2.19}  \bea
u_1 &=& \tfrac{1}{2}x\left[(p_1' m^2)' - \tfrac{1}{3}x^2 p_{1}''\right],\\
w_1 &=& \tfrac{1}{2} p_1' (x^2 - m^2),
\eea \ees
where $'$ denotes differentiation with respect to $z$ and recalling that $m=m(z)$. We derive an equation for the pressure in $\Omega_1$ by applying the boundary condition \eqref{2.17}:
\be\label{2.20}
\tfrac{1}{3}\left(p_1'm^3 \right)' = \kappa\left(p_1 - p_{2}\right).
\ee
The inlet pressure and closed end at $z=1$ from \eqref{2.13a} and \eqref{2.14a} provide the two necessary boundary conditions for the pressure in $\Omega_1$:
\bes \label{2.21}\bea 
p_1 &=& 1 \quad\hspace*{4pt} \mbox{ at } z=0,\\[0.3em]
p_1' &=& 0 \quad\hspace*{4pt} \mbox{ at } z=1.
\eea \ees

Similarly, in $\Omega_{2}$, we solve (\ref{2.12}{\color{red}\textit{b}}) and (\ref{2.12}{\color{red}\textit{c}}) using boundary conditions \eqref{2.15b} and \eqref{2.16b} to derive
\bes \label{2.22} \bea
u_{2} &=& \tfrac{1}{2}(x-1)\left[\left(p_2' (m+h-1)^2 \right)'- \tfrac{1}{3}(x-1)^2 p_{2}''\right],\\
w_{2} &=& \tfrac{1}{2} p_{2}' \left[ (x-1)^2 - (m+h-1)^2\right].
\eea \ees
The boundary condition \eqref{2.18} provides an equation for the pressure in $\Omega_2$:
\be \label{2.23}
\tfrac{1}{3}\left[p_{2}'(m+h-1)^3 \right]' = \kappa\left(p_1 - p_{2}\right).
\ee
The governing equation \eqref{2.23} is coupled to the following boundary conditions, which arise from \eqref{2.13b} and \eqref{2.14b}:
\bes \label{2.24}\bea 
p_{2} &=& 0 \quad\hspace*{4pt} \mbox{ at } z=1,\\[0.3em]
p_{2}' &=& 0 \quad\hspace*{4pt} \mbox{ at } z=0.
\eea \ees
Thus, the problem is reduced to solving \eqref{2.20}--\eqref{2.21} for $p_1$ in $\Omega_1$ and \eqref{2.23}--\eqref{2.24} for $p_{2}$ in $\Omega_{2}$. The parameters in the reduced system are the membrane permeance $\kappa$, the membrane thickness $h$, and the position of the membrane described by $a$ and $\beta$ in \eqref{2.10}.

Our goal is to understand how to maximise the flux through the filter for an applied pressure difference across the filter. As such, it is helpful to define the dimensionless flux through the system, calculated at the inlet $z=0$ as follows
\be \label{2.26}
Q = \int_{0}^m w_1|_{z=0}\,\d x = -\tfrac{1}{3} \left[p_1'm^3\right] |_{z=0}.
\ee
The system parameters are $a$, $\beta$, $\kappa$, and $h$; our goal is therefore to find the system setup that maximises $Q(a,\beta,\kappa,h)$. 
We note that the analysis of this system is much clearer in the limit of small $h$. Moreover, this limit is a regular perturbation and a finite $h$ will only slightly change the quantitative results. We consequently consider the case $h = 0$ in Section~\ref{sec:vary_a_beta}. However, as is realistic for the physical problem, we examine the quantitative changes that arise from accounting for $h>0$ in Section~\ref{sec:vary_k_h}.

\section{The effect of angling the membrane}\label{sec:vary_a_beta}
\reseteqnos{3}

In this section, we seek to understand the effect of the membrane angle on the flux through the device. We consider a setup in which we prescribe the membrane permeance $\kappa$ and thickness $h$, and study the dependency of the maximum flux achievable and the associated membrane position on the membrane angle. We consider the regular limit of a thin membrane, taking $h=0$. Taking this limit greatly simplifies the analysis, and, as previously mentioned, the results we obtain are qualitatively consistent for $h\neq 0$. 

In Section~\ref{sec:asymptotics}, we first explore the limit of a slightly angled membrane within the dimensionless domain, corresponding to $\beta \ll 1$. Recall that $\hbeta = \epsilon \beta$ so the dimensional membrane angle is always small due to the long and thin domain; the dimensionless angle is such that $\beta \in [0,1]$. In Section~\ref{sec:numerics_k1}, we then examine the full angle domain $\beta \in [0,1]$ to find the optimal position and angle associated to the maximum flux.

\subsection{Slightly angled membrane: $\beta \ll 1$}\label{sec:asymptotics}

We first consider the limit of membrane that is slightly angled away from the vertical. Since we are able to make analytical progress in this limit, this provides insight into the benefit of angling membranes. Within this limit, there are two important sub-limits, corresponding to the membrane being either far from or close to the symmetry line at $x=0$. Mathematically, these correspond to $a = O(1)$, for a centred membrane, and $a = O(\beta)$, for a membrane positioned near the corner of the domain.

\subsubsection{Membranes positioned close the centre: $a = O(1)$}\label{sec:asym_a1}

We start by considering the limit $\beta \ll 1$ with $a = O(1)$, where the membranes are slightly angled but well separated. This limit is a regular perturbation of the problem considered in \cite{herterich2017optimizing}, but imposing a constraint on the pressure difference across the filter rather than the flux. We pose the following asymptotic expansions:
\bes \label{3.1}\bea
p_1 &=& p_{10} + \beta p_{11} + O(\beta^2),\\
p_{2} &=& p_{20} + \beta p_{21} + O(\beta^2).
\eea \ees

Applying \eqref{3.1} to the governing equations \eqref{2.20}--\eqref{2.21}, boundary conditions \eqref{2.23}--\eqref{2.24}, and the membrane position $m(z)$ \eqref{2.10}, gives the following leading-order problem: 
\bes \label{3.2}\bea
a^3p_{10}'' &=& 3\kappa(p_{10}-p_{20}), \qquad p_{10}(0) = 1, \quad p_{10}'(1) = 0,\\
(a-1)^3p_{20}'' &=& 3\kappa(p_{10}-p_{20}), \qquad p_{20}(1) = 0, \quad p_{20}'(0) = 0.
\eea \ees
We solve the coupled system of ODEs \eqref{3.2} analytically, to yield
\bes\label{3.3}\bea
p_{10} &=& \frac{1}{N}\left[a^3 (1-a)^3 + a^6 \cosh(M) + (1-a)^6 \cosh(M(z-1))\right. \nonumber\\
& & \qquad + \left. a^3 (1-a)^3 \cosh(M z) - a^3 (1-a)^3 M(z-1)\sinh(M) \right],\\
p_{20} &=& \frac{1}{N}\left[a^3 (1-a)^3 + a^6 \cosh(M) - a^3 (1-a)^3 \cosh(M(z-1))\right. \nonumber\\
& & \qquad -\left. a^6 \cosh(M z) - a^3 (1-a)^3 M (z-1)\sinh(M )\right],
\eea\ees
where
\bea 
\kappa_1 &=& \frac{3\kappa}{a^3}, \quad
\kappa_2=  \frac{3\kappa}{(1-a)^3}, \quad
M =\sqrt{\kappa_1+\kappa_2},\nonumber\\
N &=& \left[(a^6 + (1-a)^6\right] \cosh(M) + a^3 (1-a)^3(2 + M \sinh(M)). \label{3.4}
\eea
Since this limit is a regular perturbation around $\beta =0$, \eqref{3.3} exactly solves the full problem described by \eqref{2.20}--\eqref{2.21} and \eqref{2.23}--\eqref{2.24} for a vertical membrane.

At $O(\beta$), we obtain the following system of ODEs and boundary conditions:
 \bes \label{3.5}\bea
\left[a^3p_{11}' + 3 a^2(\tfrac{1}{2}-z)p_{10}'\right]' &=& 3\kappa(p_{11}-p_{21}), \quad p_{11}(0) = 0, \quad p_{11}'(1) = 0,\nonumber\\
& & \\
\left[(a-1)^3p_{21}'+ 3(a-1)^2 (\tfrac{1}{2}-z)p_{20}'\right]' &=& 3\kappa(p_{11}-p_{21}), \quad p_{21}(1) = 0, \quad p_{21}'(0) = 0.\nonumber\\
& &
\eea \ees
Using variation of parameters, we derive the following analytic solutions to the system \eqref{3.5}:
\bes  \label{3.6}\bea
p_{11} &=& \alpha_1(z) + \alpha_2(z) z + \kappa_1 \alpha_3(z) \cosh(Mz) + \kappa_1 \alpha_4(z) \cosh(M(z-1)),\\
p_{21} &=& \alpha_1(z) + \alpha_2(z) z - \kappa_2 \alpha_3(z) \cosh(Mz) - \kappa_2 \alpha_4(z) \cosh(M(z-1)),
\eea \ees
where $\alpha_i(z) = f_i(z) + c_i$ for $i = 1,2,3,4$, with
\bes  \label{3.7}\bea
f_1(z) &=& -\frac{1}{M^2}\int_0^z s(\kappa_2 F_1(s) + \kappa_1 F_2(s))\,\d s,\\
f_2(z) &=& -\frac{1}{M^2}\int_z^1 (\kappa_2 F_1(s) + \kappa_1 F_2(s))\,\d s,\\
f_3(z) &=& \frac{\mbox{csch}(M)}{M^3}\int_z^1 \cosh(M(s-1))(F_2(s) - F_1(s))\,\d s,\\
f_4(z) &=& \frac{\mbox{csch}(M)}{M^3}\int_0^z  \cosh(M s)(F_2(s) - F_1(s))\,\d s,
\eea \ees
functions $F_1$ and $F_2$ defined as
\bes \label{3.8} \bea
F_1(z) &=& - \frac{3}{a}\left[\left(\tfrac{1}{2}-z\right)p_{10}'\right]',\\
F_2(z) &=& \frac{3}{1-a}\left[\left(\tfrac{1}{2}-z\right)p_{20}'\right]',
\eea \ees
and constants $c_1, c_2, c_3, c_4$ that satisfy
\bes \label{3.9}\bea
0 &=& c_1 + \kappa_1 (f_3(0) + c_3) + c_4 \kappa_1 \cosh(M),\\[0.25em]
0 &=&  c_2 + \kappa_1 c_3 M \sinh(M),\\[0.25em]
0 &=&  f_1(1) + c_1 + c_2 - \kappa_2 c_3 \cosh(M) - \kappa_2 (c_4 + f_4(1)),\\[0.25em]
0 &=&  f_2(0) + c_2 + \kappa_2 c_4 M \sinh(M).
\eea \ees
The system \eqref{3.9} is derived from applying the boundary conditions in \eqref{3.5} to \eqref{3.6}.

Since $p_{10}$ and $p_{20}$ are given explicitly in \eqref{3.3} and the system \eqref{3.9} is linear, the solution \eqref{3.6} represents an analytic solution to the $O(\beta$) problem. Moreover, we note that the integrals in \eqref{3.7} can be evaluated explicitly, so our solution \eqref{3.6} can be re-written in closed form. However, the full expression is unwieldy, and does not provide any additional physical insight, so we do not present it herein. Hence, we have derived asymptotic solutions to the system \eqref{2.20}--\eqref{2.21} and \eqref{2.23}--\eqref{2.24} that are accurate up to $O(\beta^2)$.\footnote{Note that in \eqref{2.10} we have chosen to describe the membrane position in terms of $a$ and $\beta$. Alternatively, one may have reasonably chosen to use $\delta_1$ and $\beta$ for which $m(z) = \delta_1 + \beta(1-z)$ (figure~\ref{fig1b}). The choice we made, however, minimises the $O(\beta^2)$ error in the subsequent asymptotic solution.}

Using the asymptotic expansion \eqref{3.1} we can derive an expansion for the flux from \eqref{2.26}:
\be \label{3.9a}
Q = Q_0+\beta Q_1 + O(\beta^2),\ee
for which we have an explicit analytic form using \eqref{3.3} and \eqref{3.6}. Importantly, this procedure results in $Q_1 >0$. Therefore, tilting a vertical membrane (so that $\beta >0$), will always improve the flux. This can be understood physically by noting that tilting elongates the membrane, increasing its surface area and, consequently, providing an easier transport route through the domain.

\subsubsection{Membranes positioned close the corner: $a = O(\beta)$}\label{sec:asym_asmall}

We now consider the limit where $\beta \ll 1$ and $a = O(\beta)$, where the membrane is still slightly angled but is now positioned close to the corner of the domain (\ie pairwise close). This regime is not a sub-limit of that studied in Section~\ref{sec:asym_a1} above, but a distinct limit in of itself. This can be seen mathematically by noting that the asymptotic results in Section~\ref{sec:asym_a1} may switch asymptotic orders as $a\to 0$. Here, the presence of an apparent corner in the domain is important. In this regime we can rewrite the membrane position \eqref{2.10} as
\be \label{3.10}
m(z) = \beta\left(A + \tfrac{1}{2}-z\right),
\ee
where $A = a/\beta$ and simply consider the single limit $\beta \to 0$ with $A=O(1)$. Using \eqref{3.10}, the governing equations \eqref{2.20}--\eqref{2.21} and boundary conditions \eqref{2.23}--\eqref{2.24} become
\bes \label{3.11} \bea
\beta^3 \left[ p_1'(A+\tfrac{1}{2}-z)^3\right]' &=& 3 \kappa (p_1-p_2),\quad p_{1}(0) = 1, \quad p_{1}'(1) = 0,\label{3.11a} \\
\left[ p_2'(\beta(A+\tfrac{1}{2}-z)-1)^3\right]' &=& 3 \kappa (p_1-p_2), \quad p_{2}(1) = 0, \quad p_{2}'(0) = 0.
\eea \ees
In the limit of $\beta \to 0$, the majority of the pressure drop across the system occurs near $z=0$. That is, there is a boundary layer of width $O(\beta^{3/2})$, where $p_1$ drops from $1$ to being of $O(\beta^{3/2})$, and $p_1=  O(\beta^{3/2})$ away from this boundary layer. Moreover, $p_2 =O(\beta^{3/2})$ everywhere in $\Omega_2$. To derive the solution in this boundary layer we introduce the boundary layer variable
\be \label{3.12}
z = \beta ^{3/2} Z.
\ee
which scales the system of ODEs \eqref{3.11} to obtain the leading-order boundary layer problem
\bes \label{3.13} \bea
(A+\tfrac{1}{2})^3\dfrac{\d^2 p_1}{\d Z^2} &=& 3 \kappa (p_1-p_2),\qquad p_{1}(0) = 1, \quad p_{1}'(\infty) = 0,\\
- \dfrac{\d^2 p_2}{\d Z^2} &=& 0, \hspace*{59pt} p_{2}(\infty) = 0, \hspace*{14pt} p_{2}'(0) = 0,
\eea \ees
where the conditions as $Z \to \infty$ arise from matching into the outer regions where $p_1, p_2 = O(\beta^{3/2})$.

The system \eqref{3.13} is solved by
\bes \label{3.14} \bea
p_1 & = & \exp\left[ - \left(\frac{3\kappa}{\left(A+\tfrac{1}{2}\right)^3}\right)^{1/2} Z\right] = \exp\left[ - \left(\frac{3\kappa}{\left(a+\tfrac{1}{2}\beta\right)^3}\right)^{1/2} z\right],\\
p_2 & = & 0,\eea \ees
and we note that the $O(\beta^{3/2})$ terms can be calculated if required. As discussed above, the solutions \eqref{3.14} show that the pressure drop in this parameter regime only occurs in $\Omega_1$ over a boundary layer of size $\beta^{3/2}$. We examine the effect of this result on the resulting flux below in Section~\ref{sec:numerics_asym}.

\subsubsection{Numerical solutions}\label{sec:numerics_asym}

\begin{figure}[ht!]
\centering
	\subfloat[][\label{fig2a}]{\includegraphics[scale=0.55]{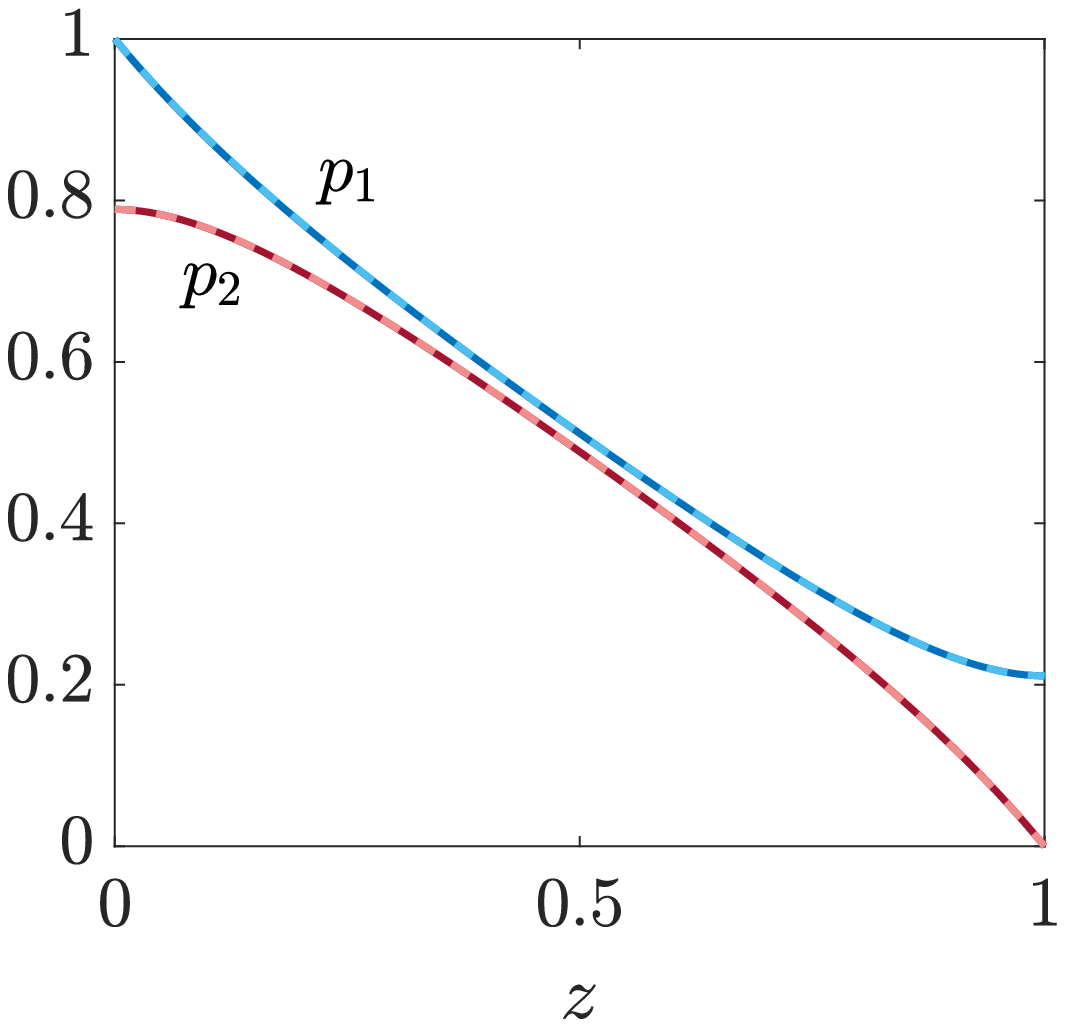}}
	\subfloat[][\label{fig2b}]{\includegraphics[scale=0.55]{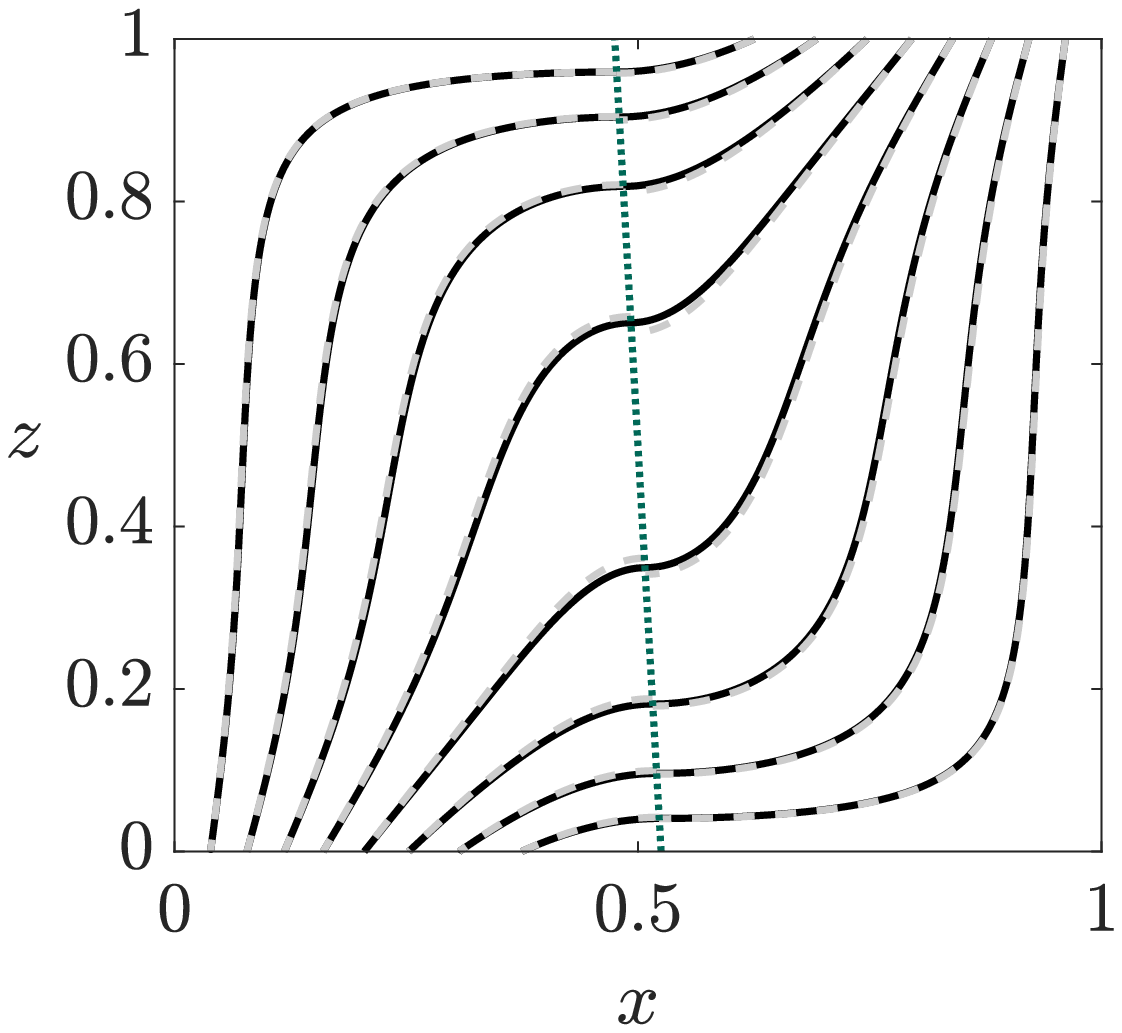}}
	\caption{Behaviour of the system governed by \eqref{2.20}--\eqref{2.21} and \eqref{2.23}--\eqref{2.24} for $\beta = 0.05$, $a = 0.5$, $\kappa=1$ and $h=0$. The pressure profiles $p_1$, $p_2$ are shown in figure~\ref{fig2a}. The membrane position (dotted green line) as given in \eqref{2.10} and streamlines are shown in figure~\ref{fig2b}. We show both numerical solutions (solid) and asymptotic solutions \eqref{3.1} (dashed) and see excellent agreement.}
	\label{fig2}
\end{figure}

In this section we present numerical results to \eqref{2.20}--\eqref{2.21} for $p_1$ in $\Omega_1$ and \eqref{2.23}--\eqref{2.24} for $p_2$ in $\Omega_2$, using $h=0$ and $\kappa=1$. We solve this system for $p_1$ and $p_2$ using MATLAB's bvp4c solver (figure \ref{fig2a}). Using the solution for the pressures, we can calculate the velocity field in each subdomain using \eqref{2.19} and \eqref{2.22}. A helpful way to visualise the qualitative behaviour of the flow through the domain is through its streamlines (figure~\ref{fig2b}). We also present the flow field predicted by our asymptotic results \eqref{3.1} for $\beta \ll 1$, $a=O(1)$, derived in Section~\ref{sec:asym_a1}, which shows excellent agreement with the numerical predictions.

We compare the flux (defined in \eqref{2.26}) predicted by the numerical results to that predicted by our asymptotic analysis for the two cases of $a= O(1)$ (in Section~\ref{sec:asym_a1}) and $a = O(\beta)$ (in Section~\ref{sec:asym_asmall}). The physically possible positions depend on the angle of the membrane: $a \in [\beta/2, 1-\beta/2]$. We note, however, that values of $a$ close to the boundaries of this interval are difficult to resolve numerically due to the presence of corners in the domain. We therefore solve for values of $a \in [ \beta/2+\delta_a, 1-\beta/2-\delta_a]$ for some small $\delta_a > 0$. To visualise the results we sample values of $\beta \ll 1$ and plot the resulting flux as a function of $a$.

\begin{figure}[ht!]
\centering
	\subfloat[][\label{fig3a}]{\includegraphics[scale=0.5]{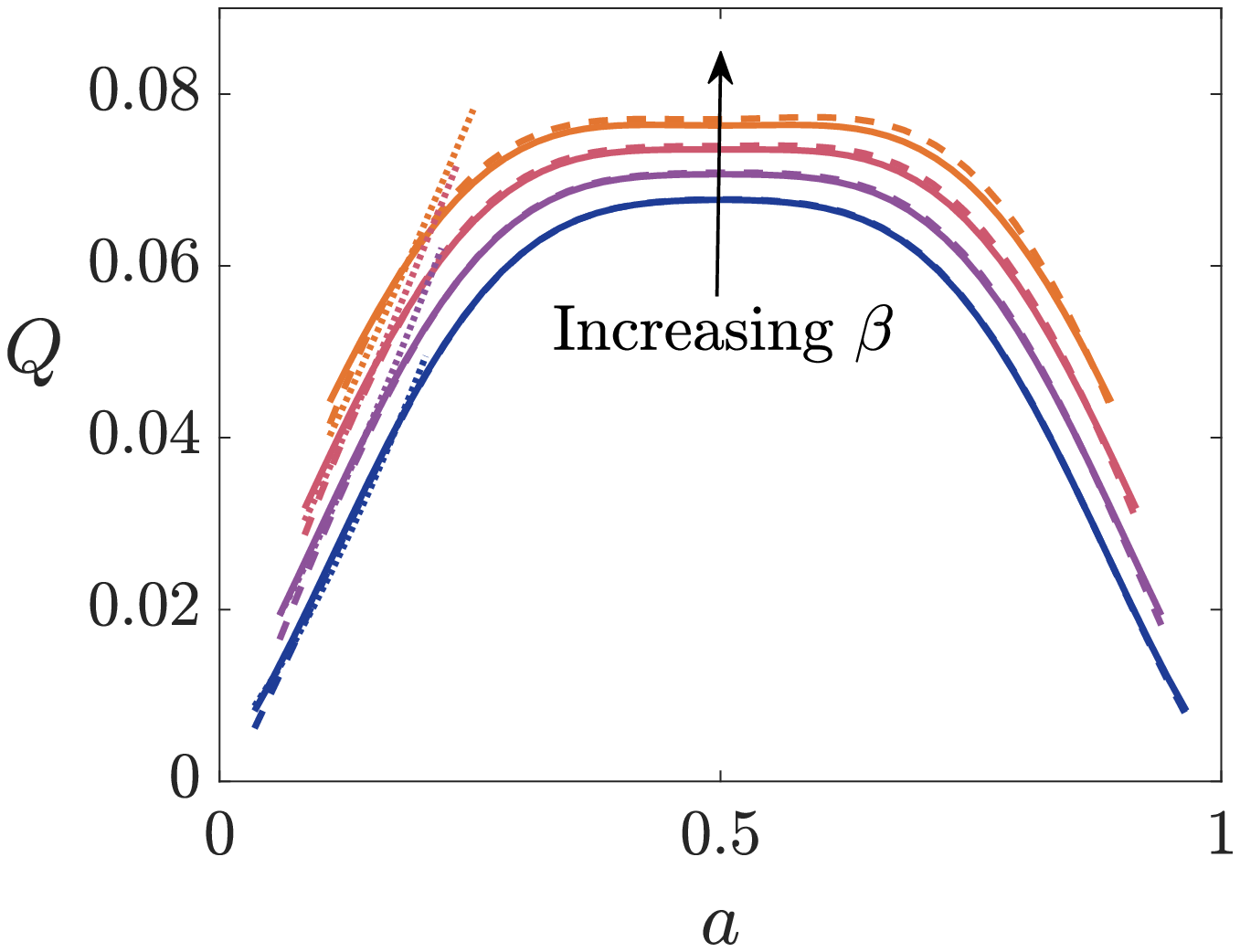}}
	\subfloat[][\label{fig3b}]{\includegraphics[scale=0.5]{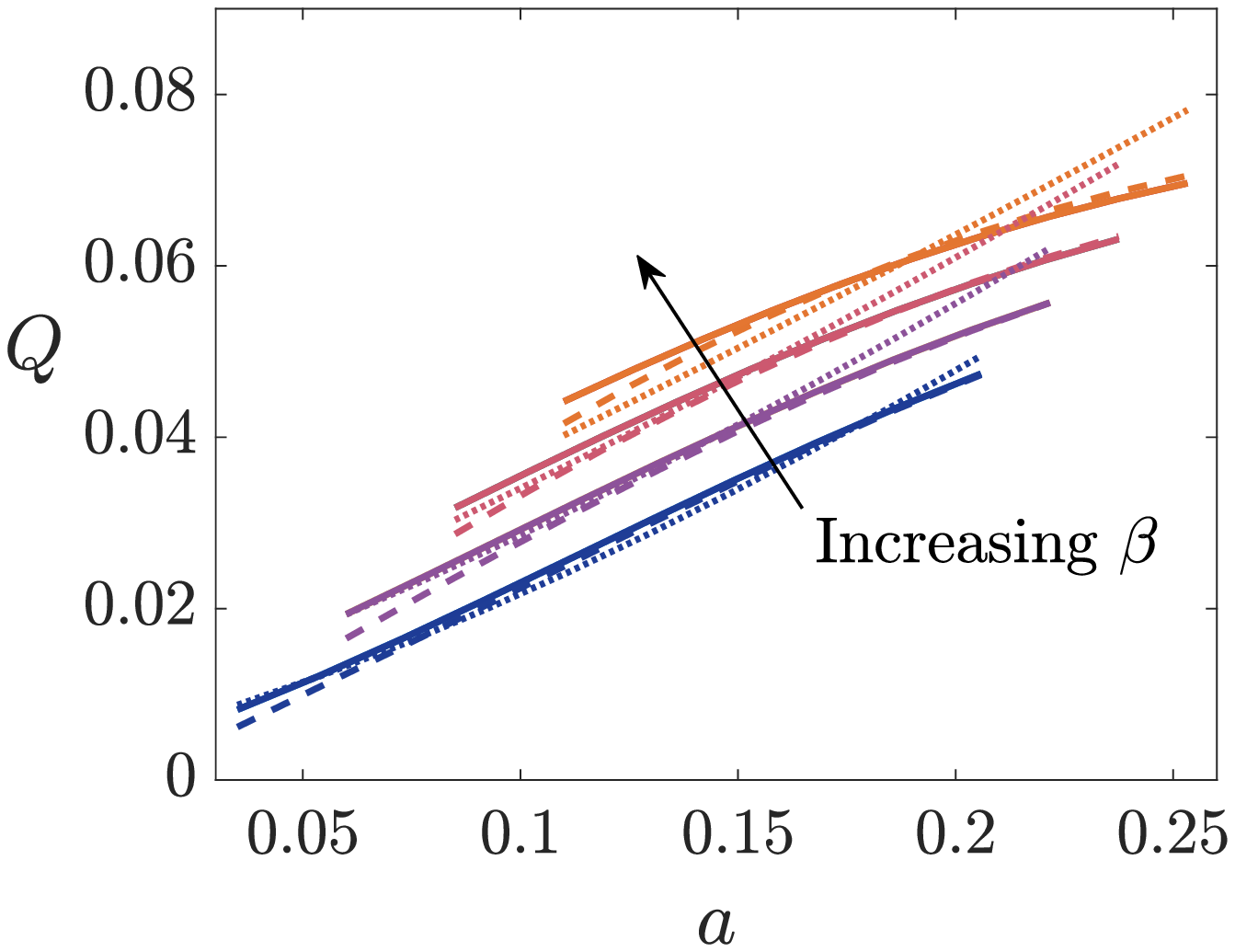}}
	\caption{Flux $Q$ given by \eqref{2.26} calculated from \eqref{2.20}--\eqref{2.21} and \eqref{2.23}--\eqref{2.24}. Numerical solutions (solid line) compared with asymptotic results for $\beta\ll 1$ and $a= O(1)$ given by \eqref{3.3} (dashed line) and for $\beta,\,a \ll 1$ given by \eqref{3.14} (dotted). Results shown for $a=O(1)$ in figure~\ref{fig3a} and for $a\ll 1$ in figure~\ref{fig3b} with $\kappa=1$, $h=0$, and $\beta = 0.05,\, 0.1,\, 0.15,\, 0.2$.}
	\label{fig3}
\end{figure}

There is excellent agreement between the numerical and analytic results for $a= O(1)$ derived in Section~\ref{sec:asym_a1} (figure~\ref{fig3a}). Importantly, we see that the flux is symmetric about $a=a^*:= 0.5$, the midpoint of the domain. Moreover, for small $\beta$, the optimal flux occurs exactly at $a=a^*$, and so it is optimal to centre the membrane in the domain. 

As we increase $\beta$ (while still keeping $\beta$ small), the curve flattens around $a=a^*$ and the flux becomes fairly insensitive to variations in $a$ near $a^*$ (figure~\ref{fig3a}). Eventually, $\partial^2Q/\partial a^2(a^*,\beta,1,0)>0$, indicating that the optimal location is no longer at $a=a^*$. However, the benefits of this optimal design over a centred design for $\beta\ll 1$ are marginal. In figure~\ref{fig3b}, we focus on the parameter regime where $a\ll 1$. As in the case when $a=O(1)$, we find that increasing the angle leads to higher fluxes. Moreover, our asymptotic results from Section~\ref{sec:asym_a1} show good agreement with the numerical results until $\beta$ gets too large. 

The analysis of this section reveals two results: first, we can increase the flux through the device for the same applied pressure difference by titling a vertical membrane; second, for a small tilt, the flux is maximised when the membrane is located  centrally, but as the membrane angle is increased the optimal location moves away from the centre and towards the corners. These observations motivate us to broaden our search to study the behaviour of the system for wider angled membranes in the following section.

\subsection{Investigating the full angle domain}\label{sec:numerics_k1}

\begin{figure}[ht!]
\centering
	\subfloat[][\label{fig4a}]{\includegraphics[scale=0.45]{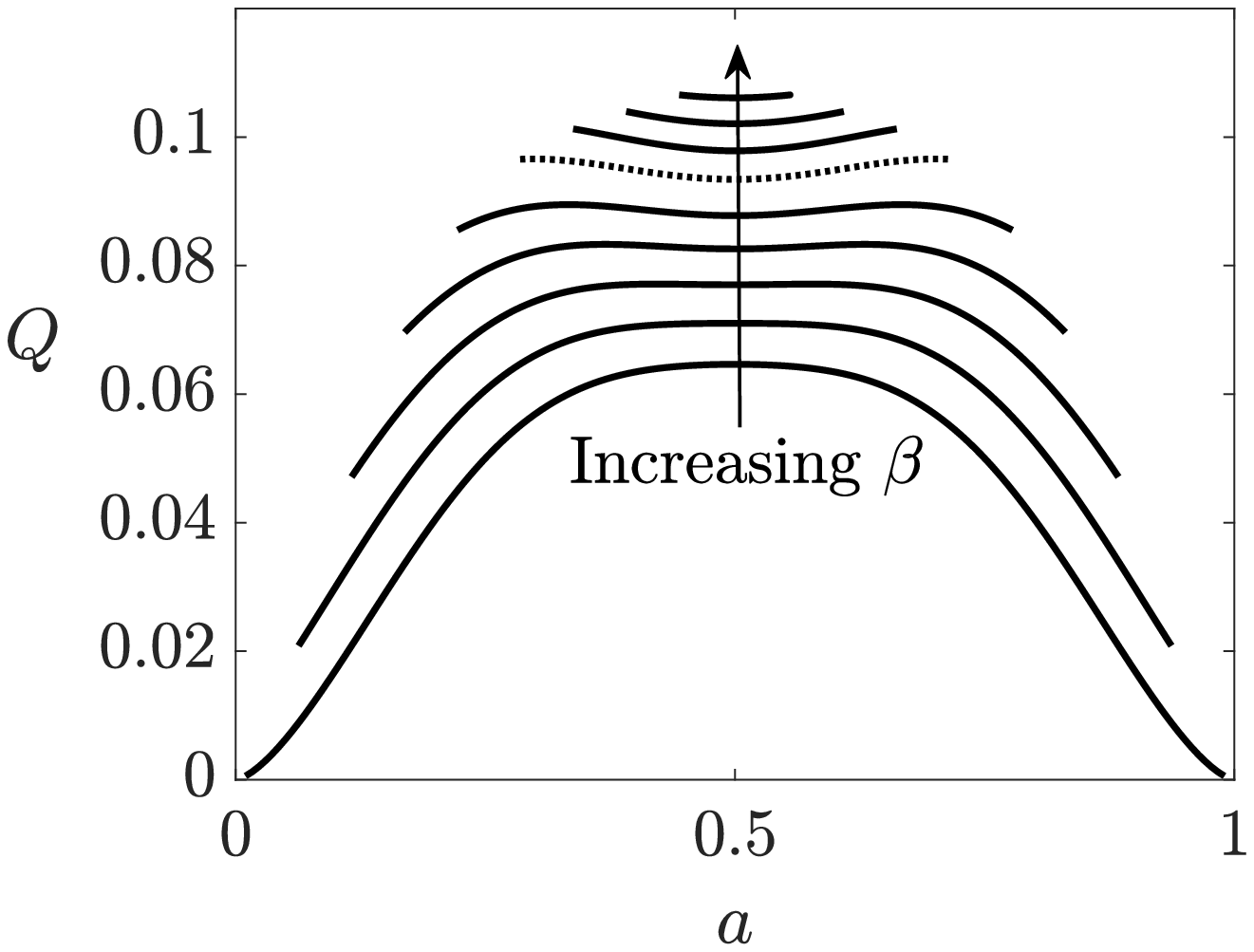}}
	\subfloat[][\label{fig4b}]{\includegraphics[scale=0.45]{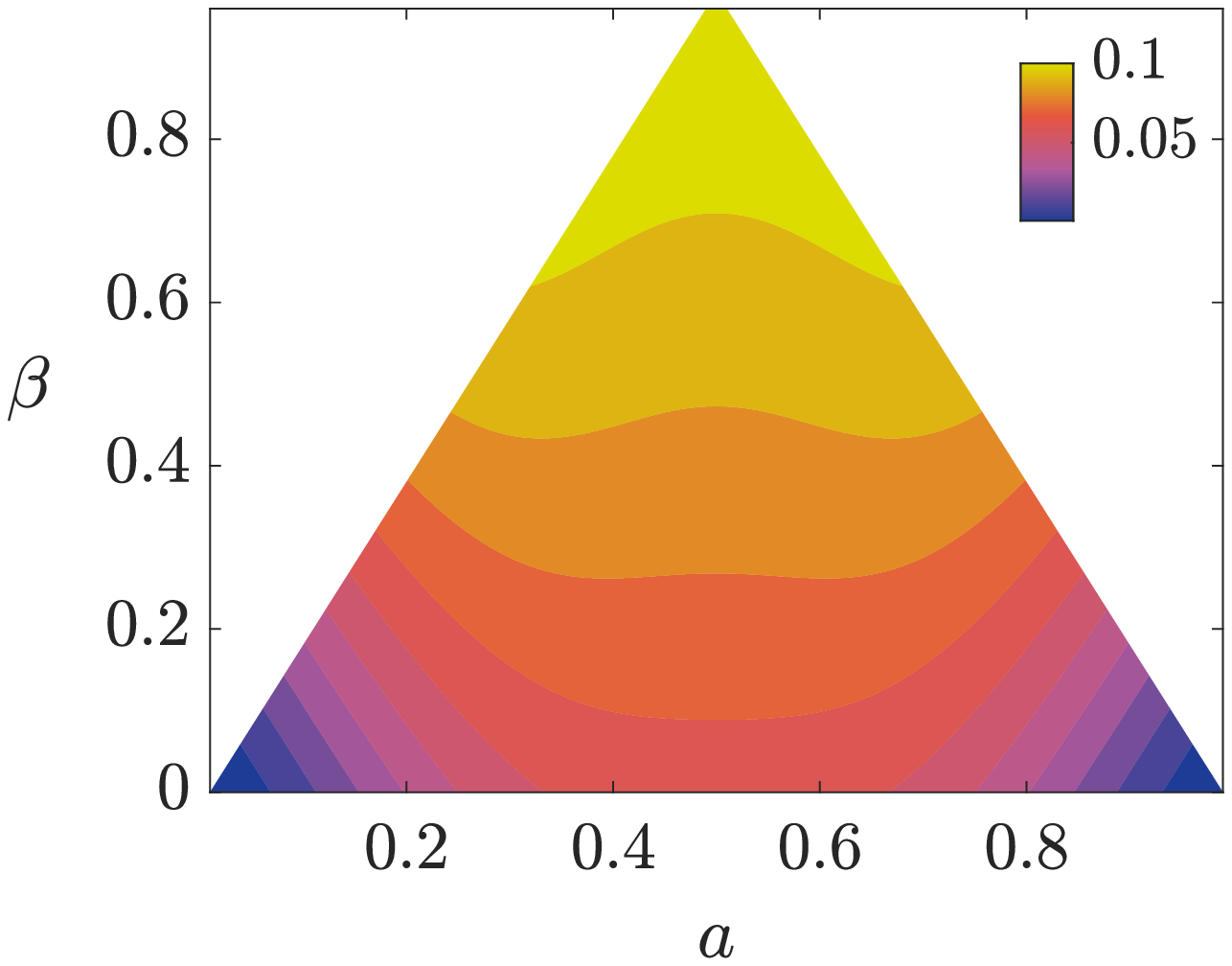}}\\
	\hspace*{-6pt}
	\subfloat[][\label{fig4c}]{\includegraphics[scale=0.45]{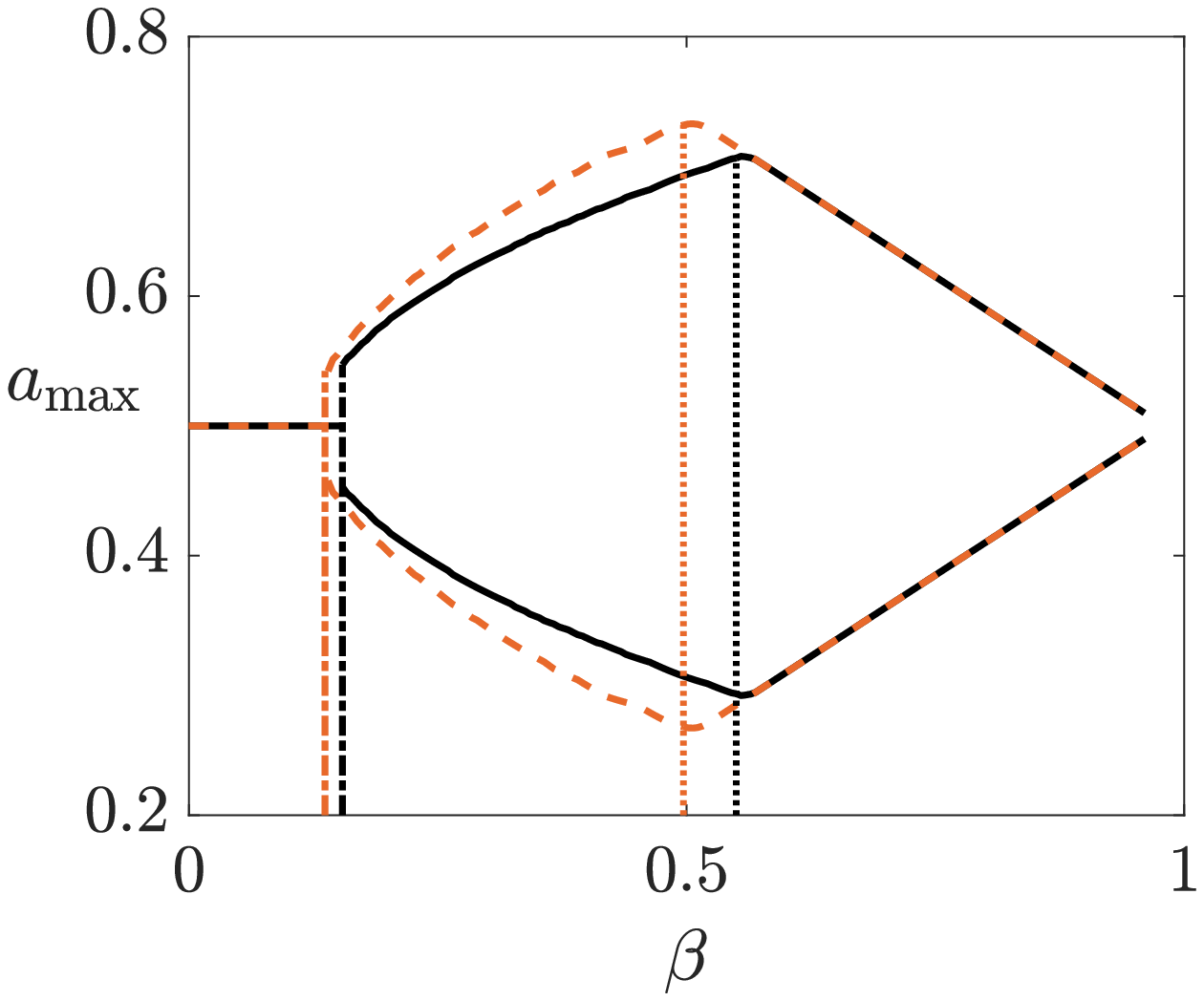}}
	\subfloat[][\label{fig4d}]{\includegraphics[scale=0.45]{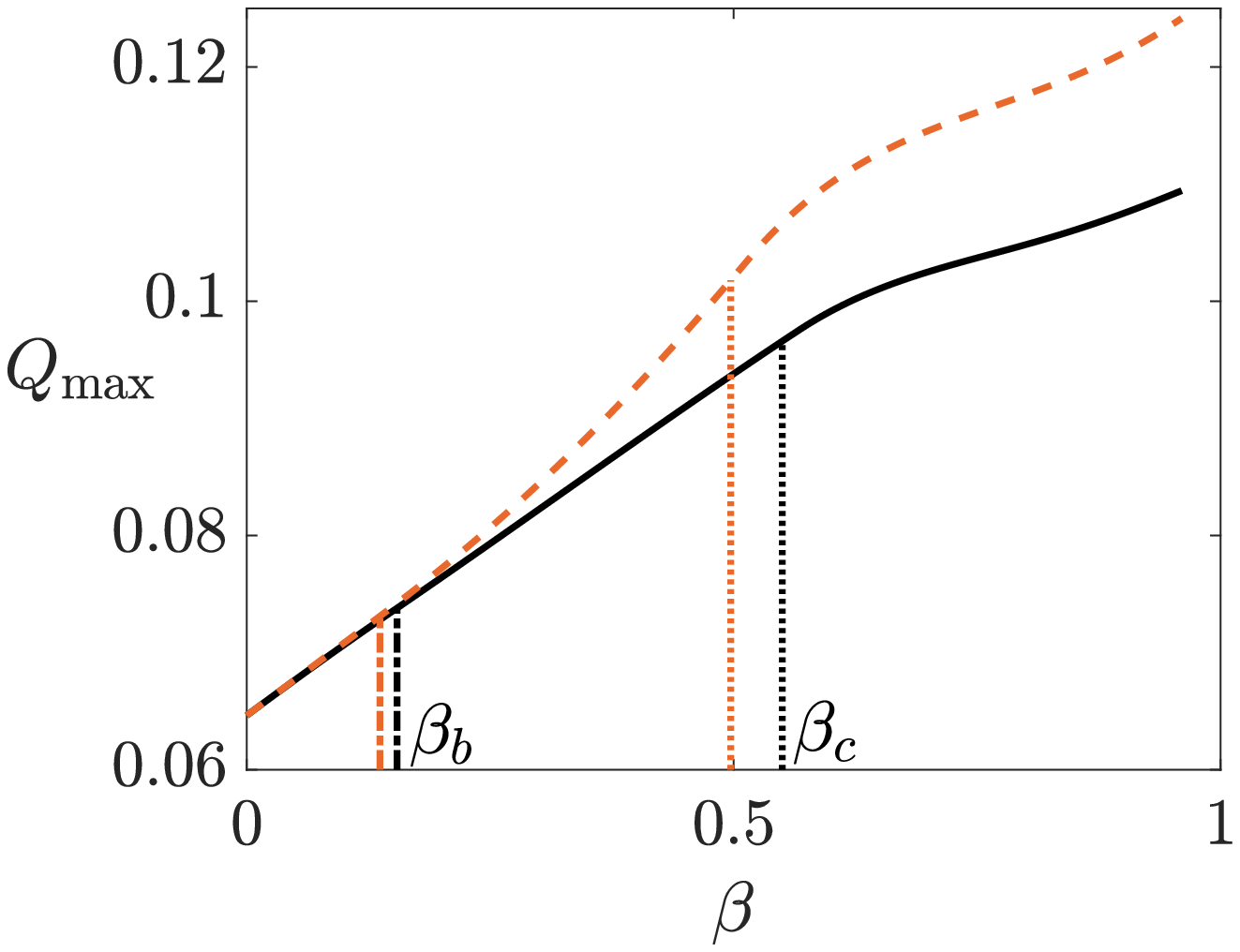}}
	\caption{Solutions to \eqref{2.20}--\eqref{2.21} and \eqref{2.23}--\eqref{2.24}. Numerical results (black, solid) and asymptotic solutions from \eqref{3.3} (orange, dashed) for \mbox{$\beta \in [0,1]$}, $a \in [\beta/2, 1-\beta/2]$ with $\kappa=1$ and $h=0$. The dash-dotted lines indicate  $\beta_b$ and the dotted lines indicate $\beta_c$.
Figure~\ref{fig4a} shows the flux $Q$ \eqref{2.26} for equally spaced $\beta \in [0, 0.87]$ inclusive, and figure~\ref{fig4b} gives the contour plot of $Q$ in $(a,\,\beta)$-space.
The optimal membrane position $a_{\max}$ corresponding to the maximum flux achievable $Q_{\max}$ \eqref{3.17} as a function of angle $\beta$ are shown in figures~\ref{fig4c} and \ref{fig4d} respectively.}
	\label{fig4}
\end{figure}

We now investigate the full angle domain, and examine the dependency of the optimal membrane position on the membrane angle. We continue to use $h=0$ and $\kappa=1$ in this section, but consider general values of these in Section~\ref{sec:vary_k_h}. We present numerical solutions to \eqref{2.20}--\eqref{2.21} for $p_1$ in $\Omega_1$ and \eqref{2.23}--\eqref{2.24} for $p_2$ in $\Omega_2$ for all values of $a$ and $\beta$ in figure~\ref{fig4}. Note that $\beta \in [0,1]$ and $a\in [\beta/2, 1-\beta/2]$.
 As mentioned in the previous section, numerical results are difficult to resolve for this system setup when corners are present in the domain. We therefore explore the solution landscape for $\beta \in [0,1-\delta_\beta]$, for some small $\delta_\beta >0$ and extrapolate our results to the full domain $\beta \in[0,1]$. Our domain for $a$ remains the same as before: $a \in [\beta/2+\delta_a, 1-\beta/2-\delta_a]$, and we extrapolate our conclusions to $a \in [\beta/2, 1-\beta/2]$.

In Section~\ref{sec:numerics_asym} we show that the flux increases with $\beta$ for small $\beta$, and this trend remains true for larger values of $\beta$ (figure~\ref{fig4a}). Moreover, in the same limit, we also show that the membrane position that maximises the flux moves away from the centre for increasing $\beta$. This trend continues in the full angle domain presented here; the optimal membrane position moves significantly off-centre as $\beta$ increases, bifurcating into two distinct off-centre optima for larger values of $\beta$ (figures~\ref{fig4a} and \ref{fig4b}). 

We define the position corresponding to the maximum achievable flux for a given angle by $a_{\max}(\beta)$. As $\beta$ increases, there is a bifurcation point $\beta = \beta_b$ at which the single, centred optimal design bifurcates into two off-centre optima, which drift further from the centre toward the corners of the domain as $\beta$ increases further (figure~\ref{fig4c}). Thus, the bifurcation point $\beta=\beta_b$ marks a fundamental change in the optimal filter design. Due to the reversibility of Stokes flow, if one off-centre optimum exists then a second optimal configuration must exist in a setup that is symmetric about $a=a^*$. 

At a second critical value $\beta = \beta_c > \beta_b$, the two optimal positions reach the corners of the physical domain, where they remain as $\beta$ increases further (figure~\ref{fig4c}). We refer to $\beta_b$ and $\beta_c$ as the bifurcation and critical points, respectively. For $\beta > \beta_b$, we define the two optimal (off-centre) positions by $a^-$ and $a^+$ (with $a^- < a^+$ and $a^- + a^+ = 1$). When $\beta > \beta_c$, and the optimal membrane position consists of one membrane end residing at a corner of the domain, and we have $a^-=\beta/2$ and $a^+ = 1-\beta/2$. 

As $\beta_b$ is the lowest value of $\beta$ at which there is a local minimum in the flux at $a=a^*$, and recalling that the flux is symmetric about $a = a^*$, $\beta_b$ is implicitly defined by the condition
\begin{align}\label{3.16}
    \dfrac{\pd^2 Q}{\pd a^2}(a^*,\beta_b,\kappa,h) = 0,
\end{align}
for a given $\kappa$ and $h$. We determine $\beta_c$ by calculating the lowest value of $\beta$ for which the two off-centre optima are such that $a^-=\beta/2$ and $a^+ = 1-\beta/2$. Finally, to understand how the flux varies as these key values of $\beta$ are crossed, it is helpful to define the maximum flux achievable over $a$-space:
\be \label{3.17}  Q_{\max}(\beta, \kappa,h) = \max_{a}\, Q(a,\beta,\kappa,h), \quad \mbox{for}\quad a\in[\beta/2,1-\beta/2],\ee
which occurs at $a = a_{\max}(\beta,\kappa,h)$.

For a membrane with permeance $\kappa=1$ and thickness $h=0$, we see that $Q_{\max}$ increases as $\beta$ increases (figure~\ref{fig4d}). There does not appear to be any change in the qualitative behaviour of $Q_{\max}$ around $\beta_b$, and this is consistent with the definition of $\beta_b$ in \eqref{3.16}. That is, as the first and second derivatives of $Q$ with respect to $a$ both vanish at this critical point, $Q$ is fairly insensitive to $a$ near $a = a^*$, $\beta = \beta_b$. However, as $\beta$ passes through $\beta_c$, there is a change in the qualitative behaviour of $Q_{\max}$. In particular, the slope of the curve decreases after this point, though we note that it does remain positive for these parameter values ($\kappa=1$, $h=0$). We will show later in Section~\ref{sec:numerics_varyk} that this is not always the case. Thus there are diminishing returns for $Q_{\max}$ in terms of increasing $\beta$ beyond $\beta_c$.

In figures \ref{fig4c} and \ref{fig4d} we also present the small-$\beta$ results derived in Section~\ref{sec:asym_a1} alongside the numerical results from the full system. The asymptotic results agree very well with the numerical results for $a_{\max}$ and $Q_{\max}$ up to $\beta \approx 0.2$, and do predict the general trend thereafter, including the presence and effect of $\beta_b$ and $\beta_c$.

The results presented in this section show that angling the membrane away from vertical can significantly increase the flux through a direct-flow device for the same applied pressure difference. Specifically, for a membrane with $\kappa=1$ and $h=0$, angling the membrane away from vertical can provide a 40\% increase in the maximum achievable flux (figure~\ref{fig4d}). Moreover, we found that while the optimal position for slightly tilted angles (with small $\beta$) is in the centre of the domain with $a=a^*$, the optimal position for increasing membrane angles bifurcates into two off-centre positions that move to the corner of the domain. Finally, the analysis of this section reveals the optimal setup that maximises the flux through the membrane with $\kappa=1$ and $h=0$ to be a membrane centred and diagonal across the full domain (with $a=a^*$ and $\beta = 1$). We now turn to the wider aim of seeking the maximum achievable flux for a membrane of finite thickness and varying permeance by varying both the membrane position $a$ and the angle $\beta$.

\section{Optimal position for a membrane of fixed properties}\label{sec:vary_k_h}
\reseteqnos{4}

In this section we address the key motivating question: for a membrane of specified permeance and thickness, how should one position and angle the membrane to maximise the flux? In Section~\ref{sec:numerics_varyk}, we retain our simplification of $h=0$ and study membranes with varying permeance. In Section~\ref{sec:numerics_varyh} we relax this condition and examine the full problem with $h>0$.

\subsection{Vanishingly thin membranes: $h= 0$}\label{sec:numerics_varyk}

\begin{figure}[ht!]
\centering
	\includegraphics[scale=0.55]{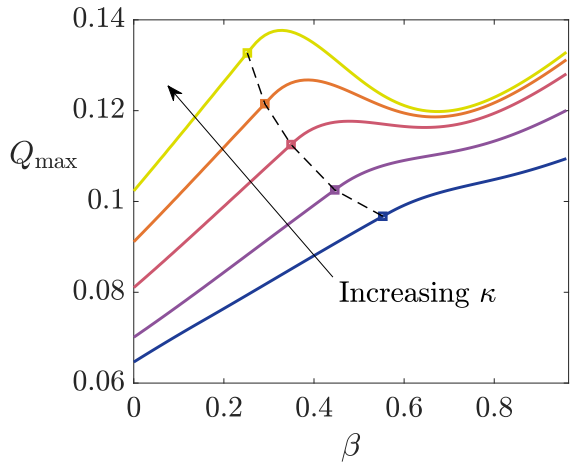}
	\caption{Maximum flux achievable $Q_{\max}$ given by \eqref{3.17} calculated from \eqref{2.20}--\eqref{2.21}, \eqref{2.23}--\eqref{2.24} and \eqref{2.26}. Results shown as a function of $\beta$ with $h=0$ and $\kappa=1$, 2, 5, 10, 20. The dashed line indicates $\beta_c$.}
	\label{fig5}
\end{figure}

For a vanishingly thin membrane, we seek the optimal position and angle of a membrane of specified permeance $\kappa$. We solve \eqref{2.20}--\eqref{2.21} and \eqref{2.23}--\eqref{2.24} numerically for $h=0$, and examine the flux \eqref{2.26} as we vary $\kappa$. Recall that the dimensionless resistance is simply the reciprocal of the permeance: $R = 1/\kappa$.
Therefore, increasing the permeance decreases the resistance, and we observe the expected increase in the maximum flux achievable for a given $\beta$ (figure \ref{fig5}).

There is a quantitative change in the profiles of $Q_{\max}$ as a function of $\beta$ for increased $\kappa$ (figure~\ref{fig5}). For small $\kappa$, the flux increases monotonically with $\beta$ and the optimal flux is achieved for $\beta =  1$ where the membrane is diagonal across the domain and centralised ($a=a^*$). For large $\kappa$, however, the highest flux is achieved when $\beta < 1$. 

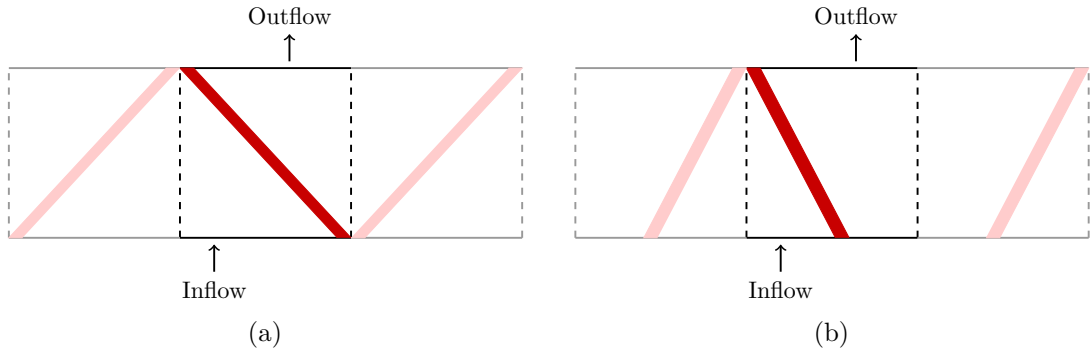
\begin{figure}
\centering
\subfloat[][\label{fig6a}]{
{\resizebox{0.46\textwidth}{!}{%
\begin{tikzpicture}[remember picture]
\draw [thick] (0,0) -- (2.5,0);
\draw [thick] (0,2.5) -- (2.5, 2.5);
\draw [thick,->] (0.5,-0.5) -- (0.5, -0.1);
\node [below] at (0.5,-0.5) {Inflow};
\draw [thick,->] (1.6,2.6) -- (1.6, 3);
\node [above] at (1.6,3) {Outflow};
\draw [thick,dashed] (0,0) -- (0, 2.5);
\draw [thick,dashed] (2.5,0) -- (2.5, 2.5);
\draw [myred, fill=myred, opacity=1] (2.5,0) -- (0.2,2.5) -- (0,2.5) -- (2.3,0);
\draw [thick, black!40] (0,0) -- (-2.5,0);
\draw [thick, black!40] (0,2.5) -- (-2.5, 2.5);
\draw [thick,dashed, black!40] (-2.5,0) -- (-2.5, 2.5);
\draw [thick, black!40] (2.5,0) -- (5,0);
\draw [thick, black!40] (2.5,2.5) -- (5, 2.5);
\draw [thick,dashed, black!40] (5,0) -- (5, 2.5);
\draw [ColorPink, fill=ColorPink] (0,2.5) -- (-0.2,2.5) -- (-2.5,0) -- (-2.3,0);
\draw [ColorPink, fill=ColorPink] (2.5,0) -- (2.7,0) -- (5,2.5) -- (4.8,2.5);
\end{tikzpicture}
}}}\quad
\subfloat[][\label{fig6b}]{
{\resizebox{0.46\textwidth}{!}{%
\begin{tikzpicture}[remember picture]
\draw [thick] (0,0) -- (2.5,0);
\draw [thick] (0,2.5) -- (2.5, 2.5);
\draw [thick,dashed] (0,0) -- (0, 2.5);
\draw [thick,dashed] (2.5,0) -- (2.5, 2.5);
\draw [thick,->] (0.5,-0.5) -- (0.5, -0.1);
\node [below] at (0.5,-0.5) {Inflow};
\draw [thick,->] (1.6,2.6) -- (1.6, 3);
\node [above] at (1.6,3) {Outflow};
\draw [myred, fill=myred, opacity=1] (1.5,0) -- (0.2,2.5) -- (0,2.5) -- (1.3,0);
\draw [thick, black!40] (0,0) -- (-2.5,0);
\draw [thick, black!40] (0,2.5) -- (-2.5, 2.5);
\draw [thick,dashed, black!40] (-2.5,0) -- (-2.5, 2.5);
\draw [thick, black!40] (2.5,0) -- (5,0);
\draw [thick, black!40] (2.5,2.5) -- (5, 2.5);
\draw [thick,dashed, black!40] (5,0) -- (5, 2.5);
\draw [ColorPink, fill=ColorPink] (0,2.5) -- (-0.2,2.5) -- (-1.5,0) -- (-1.3,0);
\draw [ColorPink, fill=ColorPink] (3.5,0) -- (3.7,0) -- (5,2.5) -- (4.8,2.5);
\end{tikzpicture}
}}}
\caption{Schematics of the two setups associated with optimal flux. The first, achieved for low permeances, is a centred membrane diagonal across full domain with $\beta=1-h$ and $a=a^*$ as shown in figure~\ref{fig6a}. The second is an angled membrane in the corner with $\beta<1$ and $a=(\beta+h)/2$ as shown in figure~\ref{fig6b}.
Note that the optimal configuration shown in figure~\ref{fig6b} has an equivalent optimal configuration for the same optimal angle $\beta<1$ with $a = 1-(\beta+h)/2$. The full concertinaed device comprises repeated modules, and we illustrate this with faded schematics for the modules neighbouring the domain of consideration.}
\label{fig6}
\end{figure}

To examine at what position $a$ this optimal flux is achieved, we consider the critical angle $\beta_c$.\footnote{One may reasonably query how the corresponding bifurcation angle $\beta_b$ varies with $\kappa$. We present the relevant results and discussion in Appendix \ref{sec:appendix}.} Recall that for $\beta \geq \beta_c$, the maximum flux is achieved when either end of the membrane is positioned in a corner of the domain. We find that the optimal flux is achieved for angles slightly larger than $\beta_c$, and thus, for large $\kappa$, the optimal membrane position involves one end being in a corner of the domain. The associated optimal angle is predicted by the model equations \eqref{2.20}--\eqref{2.21}, \eqref{2.23}--\eqref{2.24} and \eqref{2.26} through maximising $Q_{\max}$ \eqref{3.17} as shown in figure~\ref{fig5}. This result highlights the underlying physics in the problem. When optimising the flux, there is an inherent trade-off between maximising the length of the membrane and maximising the available space in one subdomain. Maximising the length of the membrane maximises the available surface area for ease of transport through the domain. Maximising the available space in either subdomain increases the transmembrane pressure drop. We find that for small $\kappa$, where the flow resistance is increased, it is more important to maximise the surface area of the membrane. Thus the optimal setup is for the membrane to be diagonal across the full domain, thereby maximising the membrane length (figure~\ref{fig6a}). For large $\kappa$, however, the membrane resistance is lower. In this case, it is more important to maximise the transmembrane pressure drop by shifting the membranes into the corners with increased available space in either subdomain (figure~\ref{fig6b}). Thus, for large $\kappa$, the optimal angle is slightly larger than $\beta_c$ which is due to the trade-off between elongating the membrane and maximising the available space in either subdomain (figure~\ref{fig5}).

\subsection{Membranes of finite thickness}\label{sec:numerics_varyh}

We have so far restricted our attention to infinitely thin membranes, but in practice membranes have a finite albeit small thickness. We now relax this assumption to consider the effect of membrane thickness on the device behaviour. Hence, the new angle domain is $\beta \in [0,1-h]$. As before, to avoid numerical complications arising from the presence of corners, we solve for $\beta \in [0, 1-h-\delta_\beta]$ and $a\in[(\beta+h)/2-\delta_a, 1- (\beta+h)/2-\delta_a)]$, for some small $\delta_\beta, \delta_a >0$, and extend our conclusions to the full domain with $\delta_\beta,\delta_a \to 0$.

In Section~\ref{sec:numerics_varyh_kfixed}, we first specify the permeance and focus on the effect of varying $h$. In Section~\ref{sec:numerics_varyh_kvary}, we then examine the combined effect of a finite thickness for varying permeance.

\subsubsection{Investigating the effect of membrane thickness}\label{sec:numerics_varyh_kfixed}

\begin{figure}[ht!]
\hspace*{-6pt}
	\subfloat[][\label{fig7a}]{\includegraphics[scale=0.5]{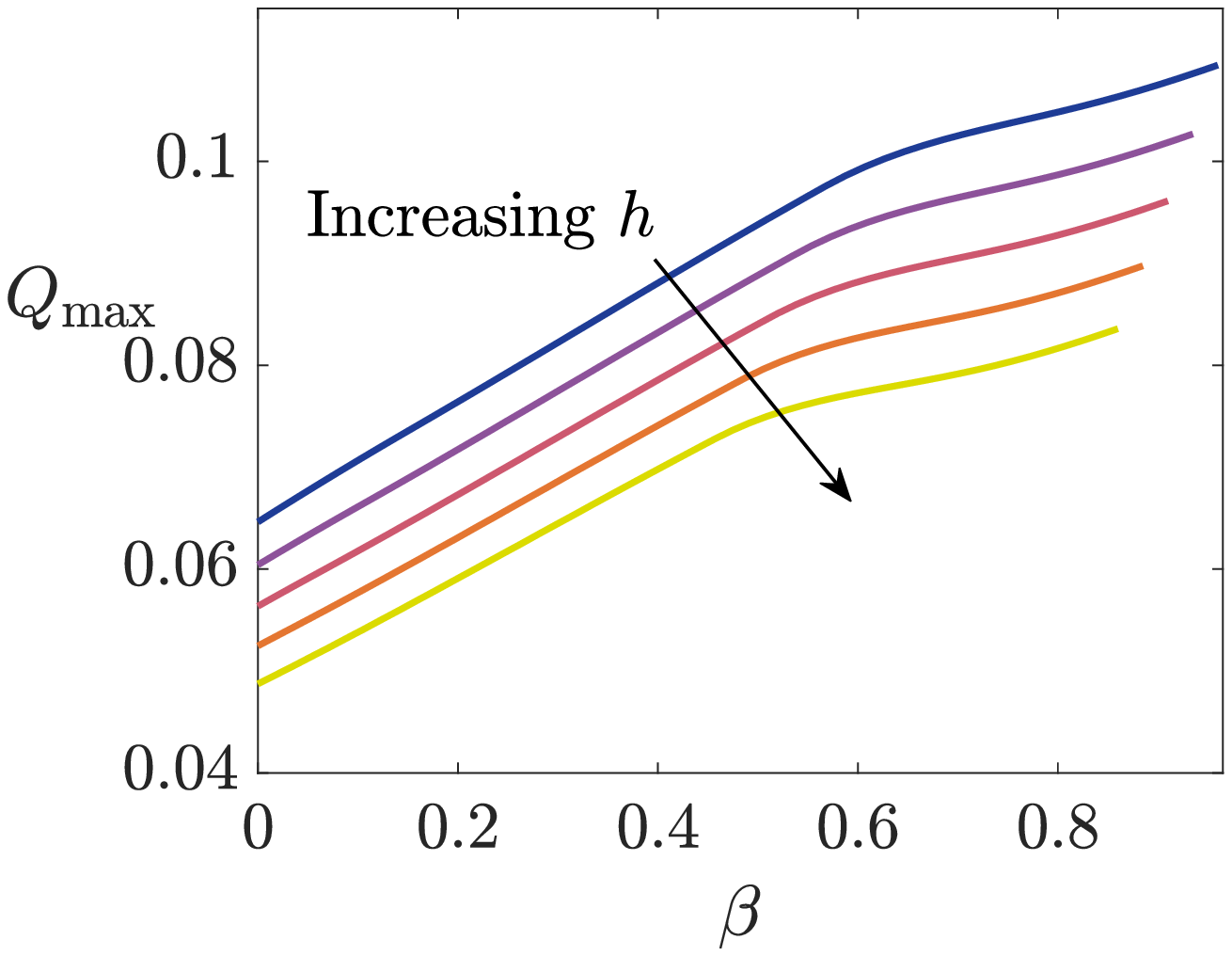}}
	\subfloat[][\label{fig7b}]{\includegraphics[scale=0.5]{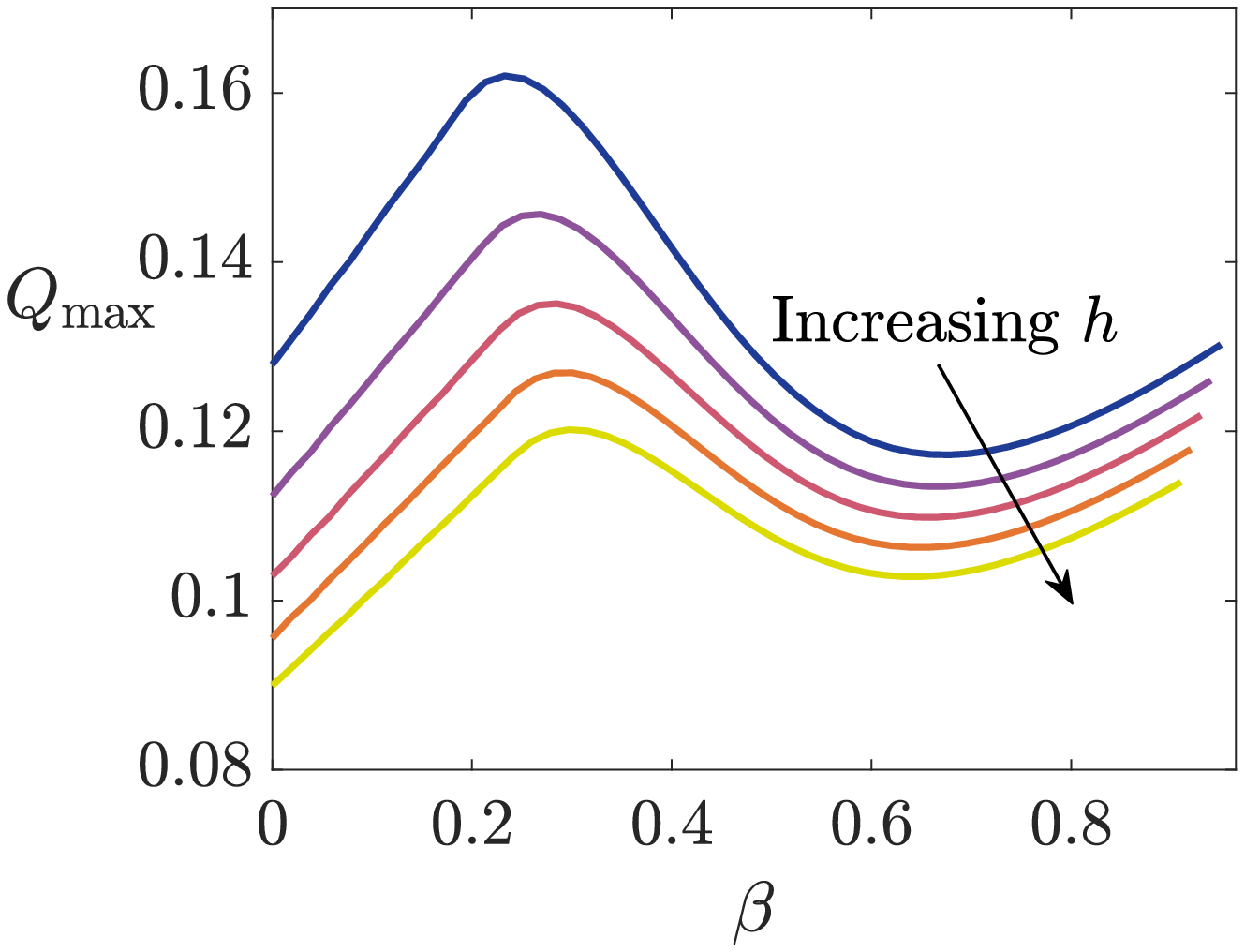}}
	\caption{Maximum flux achievable $Q_{\max}$ given by \eqref{3.17} calculated from \eqref{2.20}--\eqref{2.21}, \eqref{2.23}--\eqref{2.24} and \eqref{2.26}. 
Results shown for $h=0$, $0.025$, $0.05$, $0.075$, $0.1$ with fixed permeance $\kappa=1$ in figure \ref{fig7a}, and for $h=0.01$, $0.02$, $0.03$, $0.04$, $0.05$ with $\kappa=1/h$ (\ie fixed permeability) in figure \ref{fig7b}.}
\end{figure}

We first examine the effect of a finite membrane thickness by varying $h$ while holding $\kappa$ fixed. This corresponds to filters with different thicknesses but equal permeance, or equivalently equal net resistance. A helpful metric for comparison is $Q_{\max}(\beta,\kappa,h)$, defined in \eqref{3.17}, for varying $h$. For fixed permeance (and resistance), with $\kappa=1$, the maximum flux decreases as $h$ increases (figure~\ref{fig7a}).  We then consider the case where we vary $h$ and set $\kappa = 1/h$. Experimentally, this corresponds to comparing membranes made of the same material, so the permeability is the same, but the permeance or resistance changes accordingly with thickness. Increasing the thickness and decreasing the permeance (\ie increasing the resistance) while holding the permeability constant decreases the maximum flux (figure~\ref{fig7b}).

In both test cases ($\kappa=1$ and $\kappa=1/h$), increasing $h$ does not change the qualitative form of the behaviour. There is a clear distinction in the profile of the $Q_{\max}$ curves between figures~\ref{fig7a} and \ref{fig7b}, and this is due to increased permeance $\kappa$ in figure~\ref{fig7b}, as observed in Section~\ref{sec:numerics_varyk}.

\subsubsection{Quantifying the optimal design over experimental parameter space}
\label{sec:numerics_varyh_kvary}

Finally we explore the full parameter space to answer the question: for a membrane of given thickness and permeance, what angle and position maximise the flux through the filter? We address this through consideration of the maximum achievable flux defined by
\bea \label{4.1}
Q_{\max}^*(\kappa,h) = \max_{a,\beta} Q(a,\beta,\kappa,h)& & \mbox{for} \quad \beta \in[0, 1-h],\nonumber\\
& & \mbox{and} \quad a\in[(\beta+h)/2,1-(\beta+h)/2],
\eea
which we say occurs at $\beta = \beta_{\max}^*$ and $a=a_{\max}^*$. Note that our earlier definition of $Q_{\max}$ in \eqref{3.17} was introduced to find the optimal position associated to the maximum achievable flux for a membrane of specified angle, thickness, and permeance. The definition of $Q_{\max}^*$ in \eqref{4.1}, however, seeks the optimal angle and position for a membrane of specified thickness and permeance.

\begin{figure}[ht!]
\centering
	\subfloat[][\label{fig8a}]{\includegraphics[scale=0.5]{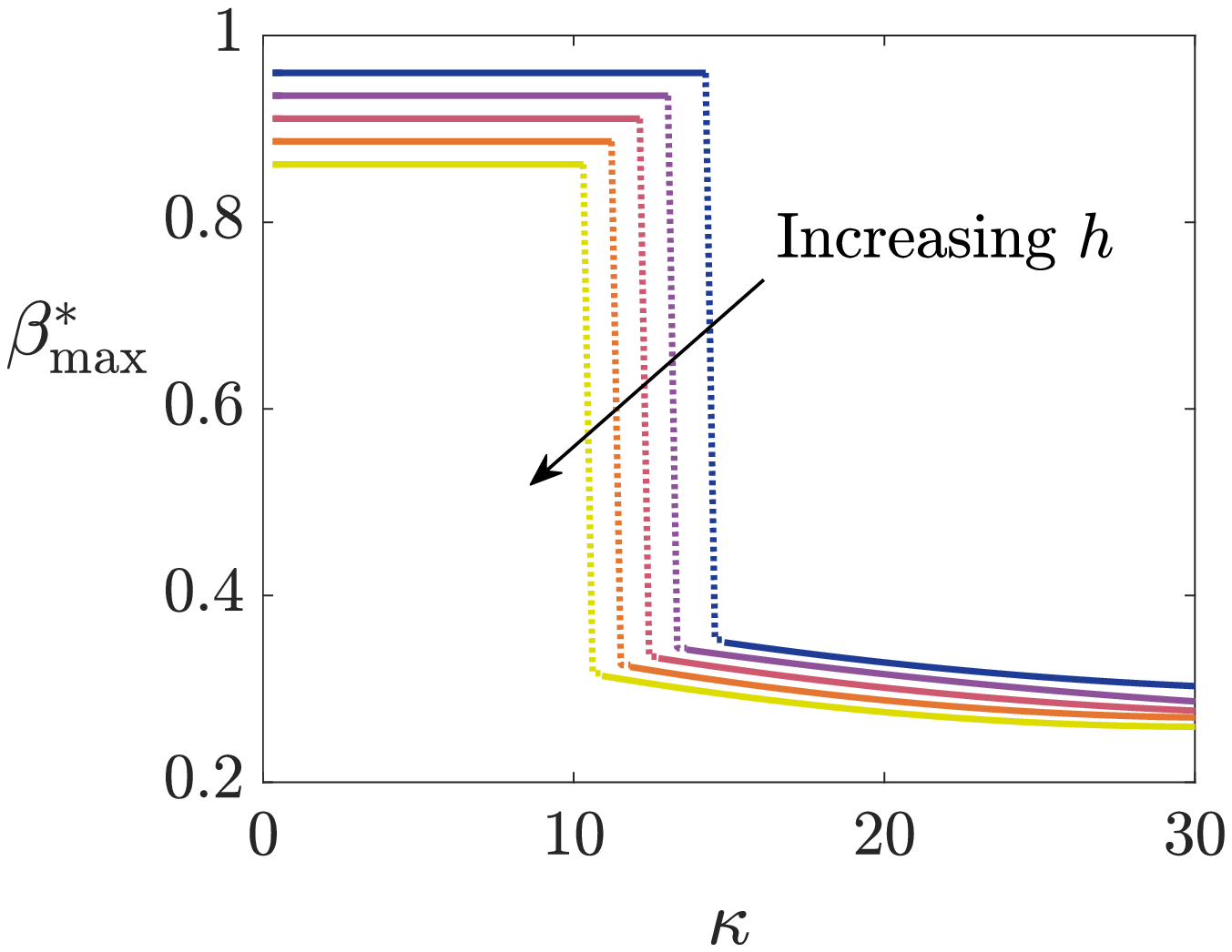}}
	\subfloat[][\label{fig8b}]{\includegraphics[scale=0.5]{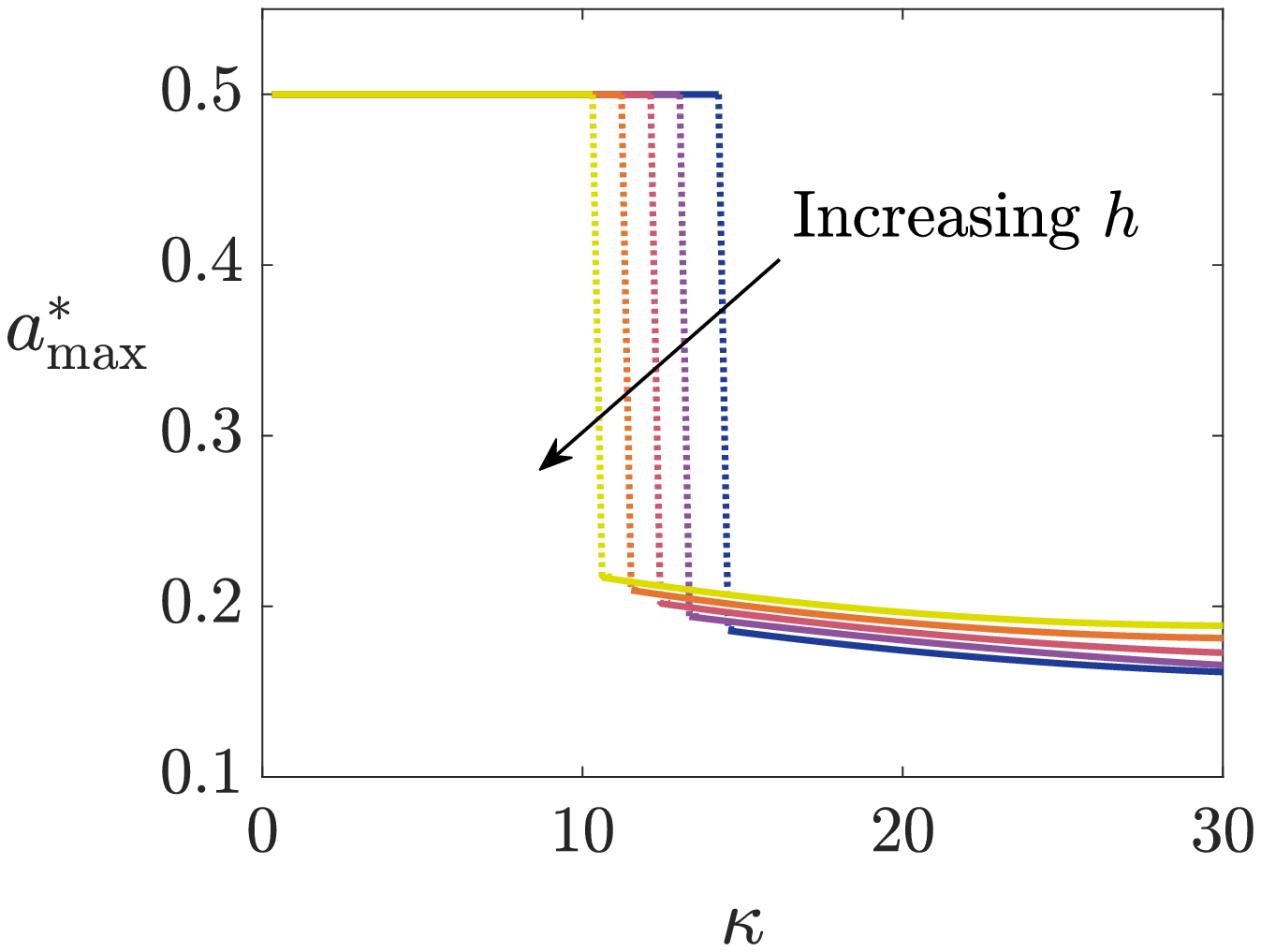}}
\caption{Angle $\beta_{\max}^*$ (figure~\ref{fig8a}) and position $a_{\max}^*$ (figure~\ref{fig8b}) associated to the maximum achievable flux $Q_{\max}^*$ defined by \eqref{4.1} calculated from \eqref{2.20}--\eqref{2.21}, \eqref{2.23}--\eqref{2.24} and \eqref{2.26}. Results shown as a function of $\kappa$ for different thickness $h=0, 0.025, 0.05, 0.075, 0.1$.}
\label{fig8}
\end{figure}

The optimal angle and position are shown in figure~\ref{fig8}. We immediately see that the profiles are discontinuous and this corresponds to the jump between the optimal positions depicted in figure~\ref{fig6}. For smaller $\kappa$, the optimal angle is $\beta_{\max}^* = 1-h$ with the corresponding optimal position at $a_{\max}^* = a^*$ (figures~\ref{fig8}).\footnote{In fact, figure~\ref{fig8} shows that $\beta_{\max}^* = 1-h-\delta_\beta$  which represents our numerically tested domain $\beta\in[0, 1-h-\delta_\beta]$. We extrapolate our conclusions to the domain $\beta \in[0,1-h]$.} This is associated to the maximum achievable flux occurring for a centred membrane diagonal across the domain (as shown in figure~\ref{fig6a}). As $\kappa$ increases, there is a critical value of $\kappa$ over which the optimal position jumps from the centre to the corner of the domain with the optimal angle occurring at $\beta_{\max}^*< 1-h$, which is determined by our model, and the associated optimal position at $a_{\max}^* = (\beta_{\max}^*+h)/2$. Note that we have, without loss of generality, presented results for the branch of solutions corresponding to $a_{\max}^* = (\beta_{\max}^*+h)/2$ only, but each off-centre optimal position is associated with a secondary optimal position symmetric about $a=a^*$, at $a_{\max}^* = 1-(\beta_{\max}^*+h)/2$ for the same $\beta_{\max}^*$. Thus, for $\kappa$ above the critical value, the maximum flux is achieved for an angled membrane in the corner of the domain (as shown in figure~\ref{fig6b}) and the angle is determined by our model. Increasing $h$ results in the critical jump occurring for smaller $\kappa$ (figure~\ref{fig8}).

\begin{figure}[ht!]
\centering
	\includegraphics[scale=0.5]{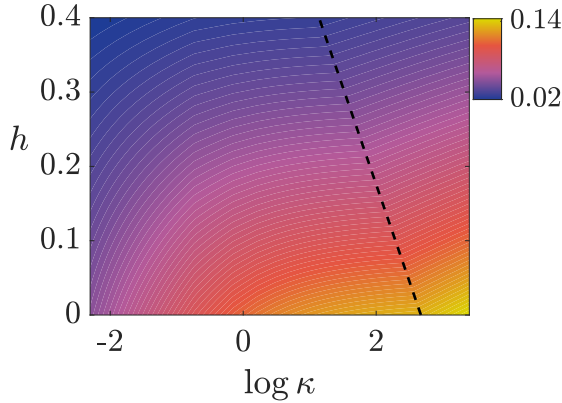}
	\caption{Maximum achievable flux $Q_{\max}^*$ in $(\kappa,\,h)$-space defined by \eqref{4.1} calculated from \eqref{2.20}--\eqref{2.21}, \eqref{2.23}--\eqref{2.24} and \eqref{2.26}. The dashed line corresponds to the critical value of $\kappa$ over where there is a discontinuous jump from centred membranes to membranes in the corners. To the left of the dashed line, the maximum flux is achieved at centred membranes diagonal across the full domain; and to the right, the maximum flux is achieved at angled membranes in the corners of the domain.}
	\label{fig9}
\end{figure}

The maximum achievable flux $Q_{\max}^*$ over a wide range of physical membrane properties is shown in figure~\ref{fig9}. We observe that the global maximum $Q_{\max}^*$ is achieved for $h \to 0$, $\kappa\to \infty$, with a vertical membrane positioned in the corner (figures~\ref{fig8} and \ref{fig9}). Physically this corresponds to maximally spaced infinitely thin membranes with zero resistance. In this case, the flux is maximised due to the pressure drop occurring almost entirely across the membrane. In practice, however, membrane design is limited in permeance and thickness. While figure~\ref{fig9} confirms that thinner, more permeable filters allow for greater flux, it provides a quantitative measure of how to design these filters for given membrane characteristics.

Importantly, for any thickness and permeance, the maximum flux is achieved either for a centred membrane diagonal across the domain ($(\kappa,\,h)$-values to the left of the dashed line in figure~\ref{fig9}) or for an angled membrane in the corner of the domain  ($(\kappa,\,h)$-values to the right of the dashed line in figure~\ref{fig9}); \ie the dashed line in figure~\ref{fig9} corresponds to the critical value of $\kappa$ over which there is a discontinuous jump in optimal design. Thus, a practitioner may use the model developed herein for a membrane of specified thickness and permeance to find the maximum achievable flux and the required angle and position that the membrane should take to achieve this flux. Specifically, if the membrane properties are to the left of the dashed line in figure~\ref{fig9}, the optimal configuration is centred and diagonal across the domain, with angle $\beta^*_{\max} = 1-h$ and position $a^*_{\max} = a^*$ (as depicted in figure~\ref{fig6a}). If the membrane properties are to the right of the dashed line in figure~\ref{fig9}, the optimal configuration is in either corner, and the angle $\beta^*_{\max}$ associated with maximum flux is provided by figure~\ref{fig8a}. The resulting optimal position is then $a^*_{\max}=(\beta^*_{\max}+h)/2$ (or equivalently $a^*_{\max}=1-(\beta^*_{\max}+h)/2$) as shown in figure~\ref{fig6b}.

\section{Conclusions}\label{sec:conc}

In this paper, we studied the steady flow through a concertinaed direct-flow filtration membrane. The aim of the work was to find the geometry of the filtration membrane within a single repeating module that maximised the flux for a given pressure drop. The physical setup facilitated a systematic asymptotic reduction of the mathematical system governing the flow. This resulted in a flow problem governed by two coupled lubrication-type equations. This system was characterised by four dimensionless parameter groupings, each representing different membrane properties; position $a$, angle $\beta$, thickness $h$, and permeance $\kappa$.

We found that angling the membrane away from vertical can greatly increase the flux through a direct-flow filtration device. For example, adjusting the device geometry for an infinitely thin membrane with fixed permeance $\kappa=1$ can increase the flux through the device by up to 40\%. Moreover, for this specific membrane, we found that for slightly tilted membranes (with a small angle) the optimal position is in the centre of the domain. Increasing the membrane angle, however, results in the optimal position bifurcating to two off-centre optima which move to the corners of the domain as the angle increases. The global optimal setup for this membrane was found to be a centred membrane diagonal across the full domain.

Extending our results to membranes of general physical characteristics, we found that there were two optimal membrane configurations. Below a critical permeance threshold, which we quantify, the membrane should be placed diagonally in a module in order to maximise membrane surface area. However, above this critical permeance threshold, the membrane should be placed in the corner of the module at an angle that depends on the system parameters, in order to maximise the transmembrane pressure drop. This discontinuity in the optimal system behaviour is summarised in figure~\ref{fig9}, which provides a guide on which behaviour is optimal for given membrane properties and resulting maximum flux.

In practice, the filtration device we modelled is used to separate fluid mixtures. As such, the membrane will block over time and change the permeance and effective shape of the membrane separating the two fluid domains. A natural extension of the work presented here is to include transient blocking dynamics and study the optimal design of the filtration device to maximise the lifespan of these filters.

The findings in this paper can be used for any direct-flow filtration device with angled membranes. Practitioners may use the model presented herein to derive the optimal configuration for a membrane of specified thickness and permeance corresponding to the maximum possible flux through the device.

\section*{Acknowledgements}

The authors would like to thank Smart Separations Ltd for presenting the initial problem, and Sotiria Tsochataridou from Smart Separations for many valuable discussions. V.\,E.\,P. is supported by the EPSRC Impact Acceleration Account Award (grant no. EP/R511742/1).
I.\,M.\,G. gratefully acknowledges support from the Royal Society through a University Research Fellowship.

\section*{Declaration of Interests}

The authors report no conflict of interest.

\bibliography{membrane}

\section*{Appendix A}\label{sec:appendix}

\begin{figure}[ht!]
\centering
\includegraphics[scale=0.5]{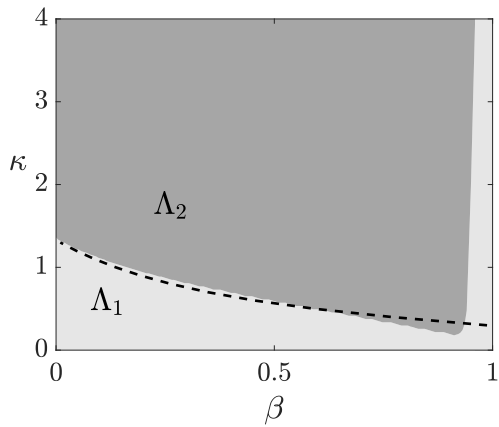}
	\caption{Phase diagram showing where one optimum membrane configuration exists in $\Lambda_1$ and where two optimum configurations exist in $\Lambda_2$ in ($\beta,\kappa$)-space. $\Lambda_2$ indicate the values $(\beta,\,\kappa)$ for which $\pd^2 Q/\pd a^2 (a,\beta,\kappa,0)>0$, where $Q$ is calculated from \eqref{2.26} using the solutions from \eqref{2.20}--\eqref{2.21} and \eqref{2.23}--\eqref{2.24}. The numerical results show the two shaded regions, and the analytical results are used to plot the black-dashed curve between them.}
	\label{fig_ap}
\end{figure}

In this appendix, we consider infinitely thin membranes with $h=0$, and examine the maximum flux achievable for membranes of varying permeance $\kappa$. Specifically, we are interested in the values of $\beta$ and $\kappa$ for which there exists a bifurcation point $\beta_b$. This is calculated by formulating the flux function $Q(a^*, \beta, \kappa,0)$ from \eqref{2.26} and deriving the domain corresponding to $\pd Q^2/\pd a^2 (a^*,\beta,\kappa,0) > 0$, for which a minimum exists at $a=a^*$ and consequently where two distinct optima co-exist. Thus we construct a phase diagram in $(\beta,\kappa)$-space (figure~\ref{fig_ap}). We denote the domain in which one finds a single optimal position by $\Lambda_1$ and that in which there are two optimal positions by $\Lambda_2$.

For permeances $\kappa \gtrsim 1.2$, the presence of the $\Lambda_2$ region indicates that there are two distinct optimal positions for the membrane angles $\beta \lesssim 0.96$ (figure~\ref{fig_ap}). For $0.96\lesssim \beta \leq 1$, there is a single optimal position. This is because larger values of $\beta$ provide tighter constraints on the membrane position within the domain, and so only one position is possible, namely the centred membrane, $a=a^*$. For permeances $\kappa\lesssim 1.2$ there is a distinctive curve separating the two regions, across which a single optimal position bifurcates into two off-centre optimal positions. Thus we conclude that $\beta_b$ exists only for small $\kappa \lesssim 1.2$.

We also show the equivalent analytic results from Section~\ref{sec:asym_a1} in figure~\ref{fig_ap}. Using the analytic expressions for the pressure in \eqref{3.1}, we are able to calculate an explicit form for the curve $\pd Q^2/\pd a^2(a^*,\beta, \kappa,0)=0$, which we show in figure~\ref{fig_ap} as a dashed line. This shows excellent agreement with the numerical results even up to intermediate values of $\beta$, despite the asymptotic results only being valid for small $\beta$. The asymptotic results do diverge significantly after $\beta \gtrsim 0.96$, since the asymptotic results do not account for the constraints in membrane position that occur at larger values of $\beta$.

\end{document}